\documentclass[twocolumn]{aastex631}

\usepackage{courier}

\DeclareRobustCommand{\ion}[2]{%
\relax\ifmmode
\ifx\testbx\f@series
{\mathbf{#1\,\mathsc{#2}}}\else
{\mathrm{#1\,\mathsc{#2}}}\fi
\else\textup{#1\,{\mdseries\textsc{#2}}}%
\fi}

\usepackage{scalerel}
\newcommand\HI{H\protect\scaleto{$I$}{1.2ex}}

\usepackage{graphicx}
\usepackage{courier}
\usepackage{amsmath} % or simply amstext
\usepackage{gensymb}
\usepackage{booktabs}

\defcitealias{2021Abdurrouf3}{paper I}

\shorttitle{Resolved scaling relations in nearby galaxies}
\shortauthors{Abdurro'uf et al.}
\begin{document}

\title{Dissecting Nearby Galaxies with \texttt{piXedfit}: II. Spatially Resolved Scaling Relations Among Stars, Dust, and Gas}

\correspondingauthor{Abdurro'uf}
\email{abdurrouf@asiaa.sinica.edu.tw}

\author[0000-0002-5258-8761]{Abdurro'uf}
\affiliation{Institute of Astronomy and Astrophysics, Academia Sinica, \\
11F of AS/NTU Astronomy-Mathematics Building, No.1, Sec. 4, Roosevelt Road, Taipei 10617, Taiwan, R.O.C.}

\author[0000-0001-7146-4687]{Yen-Ting Lin}
\affiliation{Institute of Astronomy and Astrophysics, Academia Sinica, \\
11F of AS/NTU Astronomy-Mathematics Building, No.1, Sec. 4, Roosevelt Road, Taipei 10617, Taiwan, R.O.C.}

\author[0000-0002-4189-8297]{Hiroyuki Hirashita}
\affiliation{Institute of Astronomy and Astrophysics, Academia Sinica, \\
11F of AS/NTU Astronomy-Mathematics Building, No.1, Sec. 4, Roosevelt Road, Taipei 10617, Taiwan, R.O.C.}

\author[0000-0002-8512-1404]{Takahiro Morishita}
\affiliation{IPAC, California Institute of Technology, MC 314-6, 1200 E. California Boulevard, Pasadena, CA 91125, USA}

\author[0000-0002-8224-4505]{Sandro Tacchella}
\affiliation{Kavli Institute for Cosmology, University of Cambridge, Madingley Road, Cambridge, CB3 0HA, UK}
\affiliation{Cavendish Laboratory, University of Cambridge, 19 JJ Thomson Avenue, Cambridge, CB3 0HE, UK}

\author[0000-0002-9665-0440]{Po-Feng Wu}
\affiliation{Institute of Astronomy and Astrophysics, Academia Sinica, \\
11F of AS/NTU Astronomy-Mathematics Building, No.1, Sec. 4, Roosevelt Road, Taipei 10617, Taiwan, R.O.C.}

\author[0000-0002-2651-1701]{Masayuki Akiyama}
\affiliation{Astronomical Institute, Tohoku University, Aramaki, Aoba, Sendai 980-8578, Japan}

\author[0000-0001-8416-7673]{Tsutomu T.\ Takeuchi}
\affiliation{Division of Particle and Astrophysical Science, Nagoya
University, Furo-cho, Chikusa-ku, Nagoya 464--8602, Japan}
\affiliation{The Research Center for Statistical Machine Learning, the
Institute of Statistical Mathematics, \\ 10-3 Midori-cho, Tachikawa, Tokyo
190---8562, Japan}



\begin{abstract}
We study spatially resolved scaling relations among stars, dust, and gas in ten nearby spiral galaxies. In a preceding paper \citep{2021Abdurrouf3}, we have derived spatially resolved properties of the stellar population and dust by panchromatic spectral energy distribution (SED) fitting using \verb|piXedfit|. Now, we investigate resolved star formation ($\Sigma_{\rm H_{2}}$--$\Sigma_{\rm SFR}$--$\Sigma_{*}$) and dust scaling relations. While the relations with all sub-galactic regions of the galaxies are reasonably tight ($\sigma \lesssim 0.3$ dex), we find that most of the scaling relations exhibit galaxy-to-galaxy variations in normalization and shape. Only two relations of $\Sigma_{\rm dust}$--$\Sigma_{\rm gas}$ and $\Sigma_{\rm dust}$--$\Sigma_{\rm H_{2}}$ do not show noticeable galaxy-to-galaxy variations among our sample galaxies. We further investigate correlations among the scaling relations. We find significant correlations among the normalization of the $\Sigma_{\rm H_{2}}$--$\Sigma_{\rm SFR}$--$\Sigma_{*}$ relations, which suggest that galaxies with higher levels of resolved $\text{H}_{2}$ fraction ($f_{\rm H_{2}}$) tend to have higher levels of resolved star formation efficiency (SFE) and specific star formation rate (sSFR). We also observe that galaxies with higher levels of resolved dust-to-stellar mass ratios tend to have higher levels of resolved sSFR, SFE, and $f_{\rm H_{2}}$. Moreover, we find that galaxies with higher global sSFR and less compact morphology tend to have higher levels of the resolved sSFR, SFE, and $f_{\rm H_{2}}$, which can explain the variations in the normalization of the $\Sigma_{\rm H_{2}}$--$\Sigma_{\rm SFR}$--$\Sigma_{*}$ relationships. Overall, we observe indications of the contributions of both global and local factors in governing the star formation process in galaxies. 
\end{abstract}



\section{Introduction} \label{sec:intro}

Our current understanding of galaxy formation and evolution has been largely contributed by studies of scaling relations. One of the most extensively studied scaling relations is the so-called star-forming main sequence (SFMS),  a correlation  between the global star formation rate (SFR) and stellar mass ($M_{*}$) of star-forming galaxies \citep[e.g.,][]{2004Brinchmann,2007Daddi,2007Noeske,2007Salim,2009Santini,2012Whitaker,2014Speagle}. This relation has been observed over a wide range of redshift ($0<z\lesssim 6$) with a relatively constant scatter of $\sigma\sim 0.2-0.3$ dex \citep[e.g.,][]{2012Whitaker, 2013Moustakas, 2014Speagle}. While the scatter of the SFMS relation seems to be constant, its normalization  (which reflects the specific SFR; $\text{sSFR}\equiv \text{SFR}/M_{*}$) declines with the cosmic time \citep[e.g.,][]{2012Whitaker, 2013Moustakas, 2014Speagle}. The SFMS relation indicates that the current SFR is proportional to the integrated star formation activities over the past. Although this relation has been extensively studied, its physical origin is yet to be fully understood \citep[e.g.,][]{2016Tacchella, 2019Matthee}.

Another crucial scaling relation is the Kennicutt--Schmidt ``law'' \citep[hereafter KS;][]{1959Schmidt, 1998Kennicutt}, which relates the SFR surface density ($\Sigma_{\rm SFR}$) and the gas mass surface density ($\Sigma_{\rm gas}$) with a power-law. In the pioneering studies by Kennicutt \citep[e.g,][]{1989Kennicutt, 1998Kennicutt, 1998Kennicutt_b}, this relation was found to have a super-linear slope of $\sim 1.3-1.4$. Subsequent studies have then found a tighter relation between the $\Sigma_{\rm SFR}$ and the molecular gas mass surface density ($\Sigma_{\rm H_{2}}$; that is, when the \HI{} component is removed) with a slope that is closer to unity \citep[e.g.,][]{2002Wong, 2008Bigiel, 2011Bigiel, 2012Kennicutt, 2017Utomo, 2019delosReyes}. The KS relation has also been established at high redshifts \citep[e.g.,][]{2013Genzel, 2013Freundlich}. Although the KS relation can be intuitively understood as the consequence of star formation being fueled by molecular gas, the origin of its low value of  normalization is still under debate \citep{2012Silk}.

The advancement of spatially resolved observations (both in spectroscopy and imaging) over the past couple of decades has opened up many avenues for investigations of the properties of galaxies on kiloparsec scales. Among the important findings from such observations \citep[see recent reviews by e.g.,][]{2016Cappellari, 2020Sanchez, 2020ForsterSchreiber} is that some of the major global scaling relations are preserved on kiloparsec-scale regions. Numerous studies have reported kiloparsec-scale analogs of the SFMS and KS relations, namely the spatially resolved SFMS, which is the relation between the stellar mass surface density ($\Sigma_{*}$) and $\Sigma_{\rm SFR}$ (hereafter rSFMS; e.g., \citealt{2013Sanchez, 2016CanoDiaz, 2016GonzalezDelgado, 2017Hsieh, 2017Abdurrouf, 2018Abdurrouf, 2018Pan, 2018Medling, 2020Enia, 2022Baker}), and the resolved KS relation \citep[hereafter rKS; e.g.,][]{2008Bigiel, 2011Liu, 2013Leroy, 2015Casasola, 2019Lin, 2020Morselli, 2021Ellison, 2021Pessa}, which is the relationship between $\Sigma_{\rm H_{2}}$ and $\Sigma_{\rm SFR}$. Some studies have recently revealed the relationship between the $\Sigma_{*}$ and $\Sigma_{\rm H_{2}}$ on kiloparsec scales, namely the resolved ``molecular gas main sequence'' \citep[rMGMS; e.g.,~][]{2013Wong, 2019Lin, 2020Morselli, 2021Ellison}. The rMGMS relation, which is less well studied than the rSFMS and rKS relations, suggests the importance of the stellar gravitational field in governing the local interstellar medium (ISM) conditions \citep{2019Lin}. The presence of these three resolved scaling relations indicates that the physical processes governing the global star formation activity may be controlled on kiloparsec or smaller scales.

The star formation scaling relations mentioned above only involve two main components, which are stars and gas. Another important component that is missing in those scaling relations is dust. Dust plays important roles in the physical and chemical processes of  galaxy evolution, such as: (1) shielding the gas from the interstellar radiation field (ISRF), thus allowing the gas to cool and in turn fuel star formation \citep{1999Hollenbach, 2009Krumholz, 2011Krumholz, 2011Yamasawa, 2012Glover}; and (2) catalyzing the formation of molecular gas \citep{1979Hollenbach, 2011Yamasawa}. As one of the baryonic components of galaxies, dust is expected to correlate with the other key properties, including $M_{*}$, SFR, and gas mass \citep[e.g.,][]{2017Popping, 2019Hou, 2021Triani}. 

While the global scaling relations between dust properties and other  properties associated with the stars and gas have been extensively studied \citep[e.g.,][]{2010daCunha, 2012Corbelli, 2012Cortese, 2014RemyRuyer, 2014Santini, 2014Ciesla, 2015daCunha, 2017Orellana, 2020Casasola},
relatively little attention has been paid to the corresponding scaling relations on kiloparsec scales \citep[e.g.,][]{2011Leroy, 2012Foyle, 2014Hughes}. \citet{2012Foyle} studied the spatially resolved properties of the dust, gas, and star formation in the nearby barred galaxy NGC 5236 (M83). The scaling relations they investigated include: (1) $\Sigma_{\text{H}_{2}}$--$\Sigma_{\rm dust}$ relation; (2) $\Sigma_{\rm gas}$--$\Sigma_{\rm dust}$ relation, where the gas mass is the sum of atomic and molecular hydrogen; and (3) $\Sigma_{\rm SFR}$--$\Sigma_{\rm dust}$ relation. To derive the spatially resolved dust surface density, they use far-infrared (FIR) imaging data from the \textit{Herschel Space Observatory} \citep{2010Pilbratt} and fit the pixel-wise FIR spectral energy distribution (SED) with  modified black-body models to obtain the dust mass and temperature. The SFR is derived from the $\text{H}\alpha$ and $24\mu$m maps, while the gas mass is obtained using the CO and \HI{} data.   

In this paper, which is the second in a series, we investigate the spatially resolved scaling relations on kiloparsec scales among the surface densities of stars, SFR, dust, and gas (in both atomic and molecular phases), with the goal of shedding light on the interplay among these baryonic components in galaxies and on how star formation in galaxies is regulated.  

To this end, we use multiwavelength imaging data of more than 20 bands that range from far-ultraviolet (FUV) to FIR taken from various telescopes and conduct spatially resolved SED fitting using our newly developed software \verb|piXedfit| \citep{2021Abdurrouf} to self-consistently derive the properties of the stellar population and dust. To obtain $\Sigma_{\rm H_{2}}$ and $\Sigma_{\rm HI}$ maps, we use CO and \HI{} spatially resolved survey data. Detailed descriptions on the data sets and the analysis methods we use in this paper are given in the first paper of this series \citep[][hereafter Paper I]{2021Abdurrouf3}.

In contrast to the majority of previous studies, which used two different methods for measuring the properties of the stellar population and dust (that is, dust properties are derived by fitting the FIR SED with  modified black-body models, while the SFR is obtained using simple prescriptions that involve $\text{H}_{\alpha}$ emission or photometry in ultraviolet and mid-infrared), we employ the SED fitting method that applies the energy balance principle on kiloparsec scales. This enables a robust and self-consistent measurement of the resolved properties of the stellar population and dust. As we have shown in \citetalias{2021Abdurrouf3}, our method gives more reliable SFR estimates compared to the simple prescriptions that involve total ultraviolet (UV) and infrared (IR) luminosity. The latter method is subject to the contamination by old stars in the dust heating, which can cause an overestimation of the SFR for relatively quiescent galaxies \citep[e.g.,][]{2003Hirashita, 2019Leja}.

This paper is organized as follows. We describe the data, sample galaxies, and methodology used in this work in Section~\ref{sec:data_sample}. We show ensemble scaling relations constructed from all sub-galactic regions of galaxies in Section~\ref{sec:ensemble_relations}. In Section~\ref{sec:indiv_relations}, we analyze variations of the scaling relations in individual galaxies. We further examine the correlations among the scaling relations in Section~\ref{sec:correlation_scaling0}. In Section~\ref{sec:discussion}, we discuss our results  and their implications, and put them in the context of previous studies. Finally, Section~\ref{sec:conclusion} presents the summary of this work.

Throughout this paper, we adopt cosmological parameters of $\Omega_{m}=0.3$, $\Omega_{\Lambda}=0.7$, and $H_{0}=70\text{km}\text{s}^{-1}\text{Mpc}^{-1}$, and use the AB magnitude system.  

%%%%%%%%%%%%%%%%%%%%%%%%%%%%%%%%%%%%%%%%%%
\section{Data, Sample, and Methodology} \label{sec:data_sample}

\subsection{Data} \label{sec:data}

We use three kinds of data sets in this work: broad-band imaging data that range from FUV to FIR, \HI{} 21 cm line emission maps, and CO intensity maps. We collect archival broad-band imaging data of 24 bands from various surveys, including the Galaxy Evolution Explorer \citep[GALEX;][]{2005Martin}, the Sloan Digital Sky Survey \citep[SDSS;][]{2000York}, the Two Micron All Sky Survey \citep[2MASS;][]{2006Skrutskie}, the Wide-field Infrared Survey Explorer \citep[WISE;][]{2010Wright}, the \textit{Spitzer Space Telescope} \citep{2003Gallagher}, and the \textit{Herschel Space Observatory} \citep{2010Pilbratt}. In order to obtain the surface densities of gas, including both the neutral and molecular components, % ($\Sigma_{\rm HI}$ and $\Sigma_{\rm H_{2}}$), 
we collect the \HI{} 21 cm and CO ($J=2\rightarrow 1$) emission line maps from The \HI{} Nearby Galaxy Survey \citep[THINGS\footnote{\url{https://www2.mpia-hd.mpg.de/THINGS/Overview.html}};][]{2008Walter} and the Heterodyne Receiver Array CO Line Extragalactic Survey \citep[HERACLES\footnote{\url{https://www2.mpia-hd.mpg.de/HERACLES/Overview.html}};][]{2009Leroy}, respectively. Basic information of the data sets is given in Table~\ref{tab:data_sets}. More detailed descriptions of the data sets are given in \citetalias{2021Abdurrouf3} (Sections 2.1 and 2.2 therein).

\begin{deluxetable*}{lllccc}
\tablenum{1}
\tablecaption{Data Sets Used in This Work \label{tab:data_sets}}
\tablewidth{0pt}
\tablehead{
\colhead{Data} & \colhead{Telescope/Instrument} & \colhead{Survey} & \colhead{Filter} & \colhead{PSF FWHM\textsuperscript{a}} & \colhead{Pixel Size}\\
\colhead{} & \colhead{} & \colhead{} & \colhead{} & \colhead{(arcsec)} & \colhead{(arcsec)}
}
\decimalcolnumbers
\startdata
Broad-band & GALEX & NGS\textsuperscript{b}, GII\textsuperscript{c}, AIS\textsuperscript{d} & FUV, NUV & $4.48$, $5.05$ & $1.5$ \\
Imaging & SDSS & DR12\textsuperscript{e} & $u$, $g$, $r$, $i$, $z$ & $1.5$, $1.5$, $1.5$, $1.0$, $1.0$ & $0.396$ \\
  & 2MASS & LGA\textsuperscript{f} & $J$, $H$, $K_{s}$ & $3.5$ & $1.0$ \\
  & WISE & AllWISE\textsuperscript{g} & $W1$, $W2$, $W3$, $W4$ & $5.79$, $6.37$, $6.60$, $11.89$ & $1.375$ \\
  & \textit{Spitzer}/IRAC\textsuperscript{h} & SINGS\textsuperscript{i}, SEIP\textsuperscript{j} & $3.6$, $4.5$, $5.8$, $8.0$ $\mu$m & $1.90$, $1.81$, $2.11$, $2.82$ & $1.2$/$0.6$\textsuperscript{k} \\
  & \textit{Spitzer}/MIPS\textsuperscript{l} & SINGS, LVL\textsuperscript{m} & $24$ $\mu$m & $6.43$ & $1.5$ \\
  & \textit{Herschel}/PACS\textsuperscript{n} & KINGFISH\textsuperscript{o}, VNGS\textsuperscript{p} & $70$, $100$, $160$ $\mu$m & $5.67$, $7.04$, $11.18$ & $1.4$, $1.7$, $2.85$ \\
  & \textit{Herschel}/SPIRE\textsuperscript{q} & KINGFISH, VNGS & $250$, $350$ $\mu$m & $18.15$, $24.88$ & $6.0$, $10.0$/$8.0$\textsuperscript{r} \\ 
\cline{1-6}
\HI{} 21 cm & VLA\textsuperscript{s} & THINGS\textsuperscript{t} & \nodata & $6.0$ & $1.5$ \\
\cline{1-6}
CO $J=2\rightarrow 1$ & IRAM/HERA\textsuperscript{u} & HERACLES\textsuperscript{v} & \nodata & $11.0$ & $2.0$ \\
\enddata
\tablecomments{
\textsuperscript{a} PSF FWHM information is based on \citet{2011Aniano}.
\textsuperscript{b} Nearby Galaxies Survey \citep{2003Bianchi_a,2003Bianchi_b,2004GildePaz}. 
\textsuperscript{c} Guest Investigators Survey. 
\textsuperscript{d} All-sky Imaging Survey. 
\textsuperscript{e} \citet{2015Alam}.
\textsuperscript{f} Large Galaxy Atlas \citep{2003Jarrett}.
\textsuperscript{g} \citet{2013Cutri}.
\textsuperscript{h} Infrared Array Camera \citep{2004Fazio}.
\textsuperscript{i} \textit{Spitzer} Infrared Nearby Galaxies Survey \citep{2003Kennicutt}.
\textsuperscript{j} \textit{Spitzer} Enhanced Imaging Products.
\textsuperscript{k} SINGS images have a spatial sampling of $1.2''\text{ pixel}^{-1}$, while SEIP images have a spatial sampling of $0.6''\text{ pixel}^{-1}$.
\textsuperscript{l} Multiband Imaging Photometer for \textit{Spitzer} \citep{2004Rieke}.
\textsuperscript{m} Local Volume Legacy \citep{2008Kennicutt, 2009Dale}.
\textsuperscript{n} Photodetector Array Camera and Spectrometer \citep{2010Poglitsch}. 
\textsuperscript{o} Key Insights on Nearby Galaxies: a Far-infrared Survey with \textit{Herschel} \citep{2011Kennicutt}. 
\textsuperscript{p} Very Nearby Galaxy Survey \citep{2012Bendo}.
\textsuperscript{q} Spectral and Photometric Imaging Receiver \citep{2010Griffin}.
\textsuperscript{r} KINGFISH images have a spatial sampling of $10.0''\text{ pixel}^{-1}$, while VNGS images have a spatial sampling of $8.0''\text{ pixel}^{-1}$.
\textsuperscript{s} Very Large Array.
\textsuperscript{t} The \HI{} Nearby Galaxy Survey \citep{2008Walter}.
\textsuperscript{u} Heterodyne Receiver Array \citep{2004Schuster}. 
\textsuperscript{v} Heterodyne Receiver Array CO Line Extragalactic Survey \citep{2009Leroy}.
}
\end{deluxetable*}

\subsection{Sample Galaxies} \label{sec:sample_galaxies}

We use the same sample of galaxies as in \citetalias{2021Abdurrouf3}, which consists of ten galaxies: NGC 0628 (M74), NGC 3184, NGC 3351 (M95), NGC 3627 (M66), NGC 4254 (M99), NGC 4579 (M58), NGC 4736 (M94), NGC 5055 (M63), NGC 5194 (M51a), and NGC 5457 (M101). We refer the reader to \citetalias{2021Abdurrouf3} for the description on the selection criteria for these galaxies and their basic characteristics (Table 2 therein), including the Hubble type, adopted distance, nuclear activity, integrated SFR, $M_{*}$, dust mass ($M_{\rm dust}$), neutral hydrogen mass ($M_{\rm HI}$), molecular hydrogen mass ($M_{\rm H_{2}}$), half-mass radius ($R_{e}$), and elliptical isophote fitting results that are characterized by the ellipticity ($e$) and position angle (PA). The $R_{e}$ represents the radius along the elliptical semi-major axis that covers half of the total $M_{*}$. Overall, our sample are spiral galaxies with relatively face-on configuration ($e<0.6$ and $b/a>0.4$) and have global properties that span relatively wide ranges: $M_{*}$ ($\sim 10^{9.7}-10^{11.2}$ $M_{\odot}$), SFR ($\sim 0.2-12.9$ $M_{\odot}\text{yr}^{-1}$), $M_{\rm dust}$ ($\sim 10^{7.2} - 10^{8.7}$ $M_{\odot}$), $M_{\rm H_{2}}$ ($\sim 10^{8.8} - 10^{10.7}$ $M_{\odot}$), and $M_{\rm HI}$ ($\sim 10^{8.6} - 10^{9.6}$ $M_{\odot}$). Since our sample are face-on, we do not correct the surface density quantities measured in our analysis for the inclination of the galaxies. Among the sample galaxies, two galaxies (NGC 4254 and NGC 4579) do not have \HI{} data because they are not covered in the THINGS survey. We still include these galaxies in the analysis of this paper because their $\text{H}_{2}$ information could still provide meaningful information for achieving the goals of this paper.

\subsection{Data Analysis and SED Fitting} \label{sec:data_analysis} 

A detailed description on the methods used in the analysis of our data sets is given in \citetalias{2021Abdurrouf3} (Section 3.1 therein). Here we only summarize the key steps of our data analysis, which are all carried out with \verb|piXedfit|. Basically, we combine all the FUV--FIR broad-band imaging data along with the \HI{} and CO maps, which are processed through point spread function (PSF) matching, spatial registration, and reprojection. As a consequence, all the data are brought to the spatial resolution and sampling of the SPIRE $350$ $\mu$m band. Based on the processed maps, we define the galaxy's region\footnote{This does not necessarily the whole ``region'' of the galaxy because it is often difficult to define such region. Therefore, the galaxy's region defined here simply means the region of the galaxy within which the SED fitting will be performed.} with the segmentation maps generated using the \verb|SExtractor| \citep{1996bertin}. After that, we calculate the fluxes and their uncertainties (in all bands) of all pixels within the region. Finally, we perform pixel binning to achieve a minimum signal-to-noise ($\text{S}/\text{N}$) ratio of $5$ (in all bands) for the spatially resolved SEDs. 

The \HI{} and CO maps are also processed through the same steps as above. Once we have the \HI{} and CO maps with the same spatial resolution as those of the broad-band imaging data, we calculate $\Sigma_{\rm HI}$ and $\Sigma_{\rm H_{2}}$ of pixels with the assumption of a constant CO-to-$\text{H}_{2}$ conversion factor of $\alpha_{\rm CO}=4.35\times 10^{6} M_{\odot}\text{kpc}^{-2}(\text{K}\text{ km}\text{ s}^{-1})^{-1}$ \citep[e.g.,][]{2013Bolatto}. 
Such an assumption is made in most of the studies of the resolved scaling relations of galaxies in the literature \citep[e.g.,][]{2008Bigiel, 2015Casasola, 2019Lin, 2020Morselli, 2021Ellison}. To investigate the effects of a non-constant $\alpha_{\rm CO}$ on the resolved scaling relations derived in this study, in Appendix~\ref{sec:effect_varian_alpha_CO}, we will apply a metallicity-dependent $\alpha_{\rm CO}$ and discuss the changes in the scaling relations.

Once the pixel binning is done, we perform SED fitting for each bin to derive the spatially resolved properties of the stellar population and dust, again using \verb|piXedfit|. 
The SED of the composite stellar population is calculated using the Flexible Stellar Population Synthesis \citep[\texttt{FSPS}\footnote{\url{https://github.com/cconroy20/fsps}};][]{2009Conroy} code.  The SED modeling in FSPS includes emission from the stars, nebulae, dust, and a dusty torus surrounding an active galactic nucleus (AGN). It implements the energy balance principle (i.e., the energy absorbed by dust in the ultraviolet to near-infrared is equal to the energy re-radiated in the infrared). The dust emission is based on the \citet{2007Draine} templates \citep[see][]{2017Leja}, while the modeling of the AGN dusty torus emission uses the \citet{2008Nenkova_a, 2008Nenkova_b} \texttt{CLUMPY} templates \citep[see][]{2018Leja}.  

In this work, we adopt the following settings in the SED modeling: an initial mass function of \citet{2003Chabrier}, Padova isochrones \citep{2000Girardi,2007Marigo,2008Marigo}, MILES stellar spectral library \citep{2006Sanchez-Blazquez,2011Falcon}, a parametric star formation history (SFH) in the form of a double power-law function, and the two-component dust attenuation law by \citet{2000Charlot}. For the fitting process, we use the Markov Chain Monte Carlo (MCMC) method. We only include the AGN component for fitting the SED of the spatial bin in the center of a galaxy that is either known to host an AGN or unclassified (see Table 2 in \citetalias{2021Abdurrouf3}). The list of free parameters in the SED modeling and fitting, along with the adopted priors, are given in \citetalias{2021Abdurrouf3} (Table 4 therein).  

As mentioned above, the SED fitting is applied to individual spatial bins in the galaxies, so we obtain inferred parameters at the spatial bin scales. For parameters that scale with the flux ($M_{*}$, SFR, and $M_{\rm dust}$), we further divide the values associated with a spatial bin into its member pixels in proportion to the flux in a certain band ($K_{s}$, $24$ $\mu$m, and $100$ $\mu$m bands for $M_{*}$, SFR, and $M_{\rm dust}$, respectively). Specifically, we weigh the pixel value with $w=f_{\rm p,i}/f_{\rm b,i}$, where $f_{\rm p,i}$ and $f_{\rm b,i}$ are the fluxes of the pixel and the parent bin, respectively. Such further division into the pixels is useful in comparing the spatially resolved $\Sigma_{*}$, $\Sigma_{\rm SFR}$, and $\Sigma_{\rm dust}$ with  $\Sigma_{\rm HI}$ and $\Sigma_{\rm H_{2}}$, which are already in pixel basis.   

%%%%%%%%%%%%%%%%%%%%%%%%%%%%%%%%%%%%%%%
\section{Spatially Resolved Scaling Relations with the Ensemble of all Pixel Data} \label{sec:ensemble_relations}

We start by first presenting two sets of scaling relations that are derived with the ensemble of all pixels data (from all galaxies). These scaling relations include: the relation among $\Sigma_{\rm SFR}$, $\Sigma_{*}$, and $\Sigma_{\rm H_{2}}$, hereafter referred to as the ``resolved star formation scaling relations'' (Section~\ref{sec:merged_SF_relations}), and correlations among the surface densities of dust ($\Sigma_{\rm dust}$) and other resolved properties (including gas mass, stellar mass, SFR, and sSFR), hereafter referred to as the ``resolved dust scaling relations'' (Section~\ref{sec:merged_dust_relations}). 

In the analysis throughout this paper, we only consider pixels that have $\Sigma_{\rm SFR}$,  $\Sigma_{*}$, $\Sigma_{\rm HI}$, and $\Sigma_{\rm H_{2}}$ above the following thresholds: $10^{-5}$ $M_{\odot}\text{ yr}^{-1}\text{ kpc}^{-2}$, $10^{6}$, $2.6\times 10^{6}$, and $3.0\times 10^{6} $ $M_{\odot}\text{ kpc}^{-2}$, respectively.  While the thresholds for $\Sigma_{\rm SFR}$ and $\Sigma_{*}$ are low and only give a weak constraint on the pixel selection, those for $\Sigma_{\rm HI}$ (which corresponds to a \HI{} column density of $3.2\times 10^{20}$ $\text{cm}^{-2}$) and $\Sigma_{\rm H_{2}}$ are about the average surface densities of the detection limits of the THINGS and HERACLES surveys (\citealt{2008Walter} and \citealt{2009Leroy}, respectively). These cuts are adopted because of the consideration that the low S/N pixels (that usually reside in the outskirts) can have significant influences on the slope and normalization of the scaling relations. For scaling relations that involve $\Sigma_{\rm HI}$ or $\Sigma_{\rm gas}$, we exclude the two galaxies that do not have \HI{} data (NGC 4254 and NGC 4579). Overall, the number of pixels included in the analysis of the scaling relations that do not involve $\Sigma_{\rm HI}$ or $\Sigma_{\rm gas}$ is 8363, while that number for the scaling relations that involve $\Sigma_{\rm HI}$ or $\Sigma_{\rm gas}$ is 7534. The physical size of pixels vary from galaxy to galaxy and it ranges from $0.2$ to $1.6$ kpc with a median and $84$th percentile of the pixel size distribution of $0.3$ and $0.5$ kpc, respectively.

\subsection{Resolved Star Formation Scaling Relations with All Pixel Data} \label{sec:merged_SF_relations}

\begin{figure*}
\centering
\includegraphics[width=1.0\textwidth]{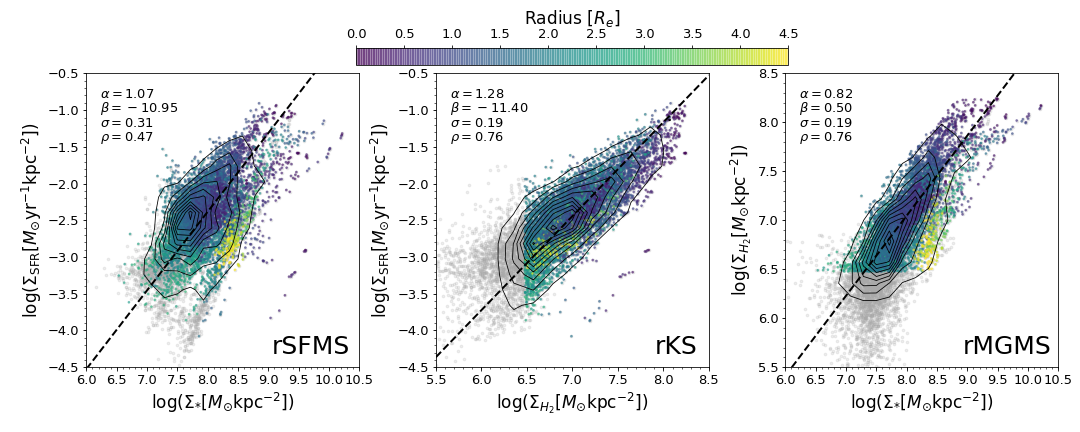}
\caption{Spatially resolved star formation scaling relations with the ensemble of all pixels: rSFMS (left panel), rKS (middle panel), and rMGMS (right panel). The contours represent the number density distribution of the pixels. The black dashed lines show the best-fit linear functions obtained from a fit using the ODR method. The slope ($\alpha$), zero-point ($\beta$), scatter ($\sigma$), and the Spearman rank-order correlation coefficient ($\rho$) of the relation are shown in the top left corner of each panel. The larger scatter of the rSFMS than the other two relations indicate that the rSFMS is likely originated from the rKS and rMGMS relations.}
\label{fig:merge_SF_bin_pixs}
\end{figure*}

We show the resolved star formation scaling relations  in Figure~\ref{fig:merge_SF_bin_pixs}. The panels from left to right show the rSFMS, rKS, and rMGMS relations, respectively. In each panel, the gray points represent the pixels that are excluded  due to the thresholds set above. The data points in the scaling relations are color coded based on the radial distance of the pixels from the galactic center measured along the elliptical semi major axis. As we can see, 
pixels of different radial distances are mixed across the locus of the three relations, in contrast to the naive expectation (based on the radial profiles) that pixels closer to the galactic center would tend to occupy the high surface density locus. This trend is likely caused by the diversity in the global properties of the sample galaxies, especially the $M_{*}$ (which spans a range of $\sim 1.5$ dex), molecular gas mass (a range of $\sim 1.8$ dex), and SFR (a range of $\sim 1.8$ dex), which define the overall level of the spatially resolved properties in individual galaxies.

The contours show the number density distributions of the pixels. For a quantitative analysis of the scaling relations, we fit each relation with a linear function of the form 
\begin{equation}
\log(\text{Y}) = \alpha \log(\text{X}) + \beta,
%\text{Y} = \alpha * \text{X} + \beta,
\label{eq:lin_func}
\end{equation} 
where $\text{Y}$ and $\text{X}$ are the physical quantities ($\Sigma_{*}$, $\Sigma_{\rm SFR}$, or $\Sigma_{\rm H_{2}}$) on the $x$- and $y$-axes of a scaling relation, respectively. The coefficients $\alpha$ and $\beta$ are the slope and zero-point, respectively. The fitting is performed using the orthogonal distance regression (ODR) method, which considers deviation (i.e., dispersion) along both $x$- and $y$-axes, in contrast to the ordinary least square (OLS) method, which only considers the deviation along the $y$-axis. Moreover, we calculate the Spearman rank-order correlation coefficient ($\rho$) in order to evaluate the strength of the correlation. The best-fit linear functions are shown in the figure as the black dashed lines. The slope, zero-point, scatter ($\sigma$), and $\rho$ of these scaling relations are summarized in Table~\ref{tab:fit_merge_SF_bin_pixs}. The uncertainties of $\alpha$ and $\beta$ are obtained using the bootstrap resampling method. Throughout this paper, we use these quantities, especially $\sigma$ and $\rho$, as an indicator for the significance of a scaling relation. Given the uncertainties of the physical quantities involved in the scaling relations, the observed scatter of the relations is always larger than the intrinsic one. Because of the complexity, we do not attempt to correct the observed scatter for the measurement errors, which include a careful accounting of systematic errors. However, we also use $\rho$ in addition to $\sigma$ for evaluating the significance of a scaling relation, which provide an independent measure.

\begin{deluxetable}{ccccc}
\tablenum{2}
\tablecaption{Results of the Fitting of a Linear Function to the Resolved Star Formation Scaling Relations \label{tab:fit_merge_SF_bin_pixs}}
\tablewidth{0pt}
\tablehead{
\colhead{Correlation} & \colhead{$\rho$} & \colhead{$\alpha$} & \colhead{$\beta$} & \colhead{$\sigma$}
}
\decimals
\startdata
rSFMS & $0.47$ & $1.07\pm 0.02$ & $-10.95\pm 0.17$ & $0.31$ \\ 
rKS & $0.76$ & $1.28\pm 0.01$ & $-11.40\pm 0.09$ & $0.19$ \\ 
rMGMS & $0.76$ & $0.82\pm 0.01$ & $0.50\pm 0.07$ & $0.19$ 
\enddata
%\tablecomments{ }
\end{deluxetable} 

Our results show that the rSFMS relation has the largest scatter ($0.31$ dex) among the three, while the rKS and rMGMS relations are similarly tight, with a scatter of $0.19$ dex. The $\rho$ values of the rKS and rMGMS relations are also similarly high ($0.76$), further indicating the significance of these relations. This result may suggest that the rKS and rMGMS relations are independent of each other, while the rSFMS relation exists as a consequence of these two relations. A similar trend (i.e.,~the larger scatter of the rSFMS compared to the rKS and rMGMS) is also observed by previous studies \citep[e.g.,][]{2019Lin, 2020Enia, 2020Morselli, 2021Ellison, 2022Baker}. \citet{2019Lin} reported scatters of $0.25$, $0.19$, and $0.20$ dex for the rSFMS, rKS, and rMGMS relations, respectively. They analyzed 14 star-forming galaxies from the ALMA-MaNGA QUEnching and STar Formation survey \citep[ALMaQUEST;][]{2020Lin}. Using the same data sets but with a larger sample (28 galaxies) that includes green-valley galaxies, \citet{2021Ellison} obtained the scatters to be $0.28$, $0.22$, and $0.21$ dex for the rSFMS, rKS, and rMGMS relations, respectively. These results are consistent with ours in that the rKS and rMGMS relations are more significant than the rSFMS.

The rMGMS relation is less studied compared to the other two relations. An early study by \citet{2011Shi} has found a correlation between  $\Sigma_{\rm gas}$ and $\Sigma_{*}$ on global (i.e., the whole galaxy) scales. \citet{2013Wong} is the first work that demonstrated the relationship between $\Sigma_{\rm H_{2}}$ and $\Sigma_{*}$ on sub-galactic scales. Recently, this relation has been studied in greater details, and has been found to be coupled with the rSFMS and rKS relations in the context of star formation on kiloparsec scales \citep{2019Lin, 2020Morselli, 2021Ellison, 2021Pessa}.

Previous studies in the literature have reported a significant diversity in the slope of the rSFMS relation, that ranges from $\sim 0.6$ to $1.4$ \citep[][]{2013Sanchez, 2016CanoDiaz, 2017Abdurrouf, 2018Abdurrouf, 2017Hsieh, 2019Lin, 2020Enia, 2021Ellison, 2021Pessa}. Such a variation can be caused by several factors, including the differences in galaxy sample selection, the methods used to measure the SFR and $M_{*}$, and the way the relations are fit. Because of these complications, we refrain from comparing the slopes of the rSFMS relation (as well as the other relations) derived here with those in the literature. Although the slope is sensitive to the above factors, it has important physical implications. The linear slope ($\alpha \sim 1$) of the ensemble rSFMS obtained here indicates a broadly constant rate of stellar mass growth (by in-situ star formation) on average over the sub-galactic regions of the sample galaxies. On the other hand, the super-linear slope of the rKS and the sub-linear slope of the rMGMS suggest that on average, the star formation efficiency ($\text{SFE}\equiv \Sigma_{\rm SFR}/\Sigma_{\rm H_{2}}$) tend to increase with increasing $\Sigma_{\rm H_{2}}$ and that the $\text{H}_{2}$ mass-to-stellar mass ratio ($f_{\rm H_{2}}\equiv\Sigma_{\rm H_{2}}/\Sigma_{*}$) tend to decrease with increasing $\Sigma_{*}$, respectively.

%%%%%%%%%%%%%%%%%%%%%%%%%%%%%%%%
\subsection{Resolved Dust Scaling Relations with All Pixel Data} \label{sec:merged_dust_relations}

Previous studies have investigated global scaling relations among the dust mass, gas mass, stellar mass, and SFR \citep[e.g.,][]{2010daCunha, 2014Santini, 2017Orellana, 2020Casasola}. On the other hand, only a small number of studies have focused on such relations on kiloparsec-scales  \citep[e.g.,][]{2012Foyle, 2014RomanDuval}. By using our data set, we examine the correlations between $\Sigma_{\rm dust}$ and the stellar population properties and the gas mass on kiloparsec scales to study the relationship between dust and star formation processes.

\begin{figure*}
\centering
\includegraphics[width=1.0\textwidth]{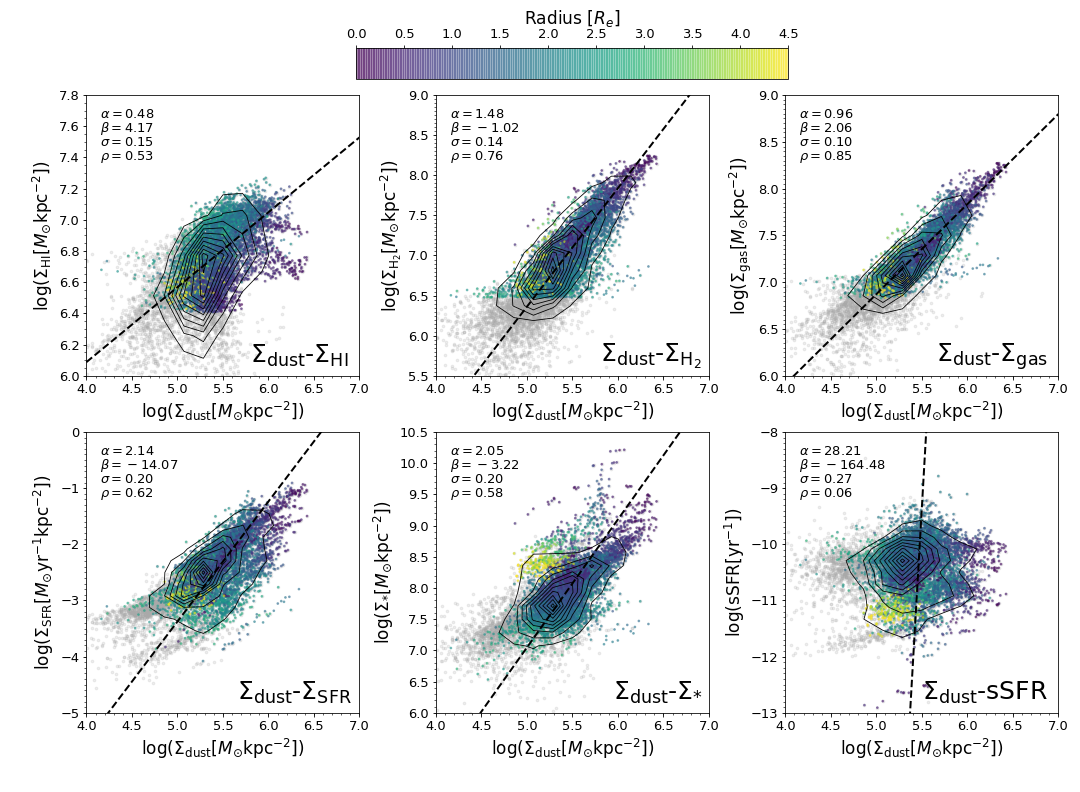}
\caption{Resolved dust scaling relations with the ensemble of all pixel data. Symbols are the same as those in Figure~\ref{fig:merge_SF_bin_pixs}. The tight $\Sigma_{\rm dust}$--$\Sigma_{\rm gas}$ and $\Sigma_{\rm dust}$--$\Sigma_{\rm H_{2}}$ relations indicate the strong coupling between the dust and gas. Moreover, dust tend to play an important role in catalyzing the formation of $\text{H}_{2}$.}  
\label{fig:edit_Mdust_MHI_MH2_bins}
\end{figure*}

Our results are shown in Figure~\ref{fig:edit_Mdust_MHI_MH2_bins}. We fit the relations with a linear function using the ODR method and summarize the results in Table~\ref{tab:fit_dust_relation}. Among the relations, we observe a tight relationship between the surface densities of dust and total gas, with a small scatter of $0.10$ dex and a high $\rho$ value of $0.85$. This relation is the tightest among the ensemble scaling relations discussed in this paper. The slope of this relation, which is close to unity ($\sigma=0.96 \pm 0.01$), indicates a nearly constant dust-to-gas mass ratio. This agrees with the broadly flat dust-to-gas mass ratio radial profile of the sample galaxies that we analyzed in \citetalias{2021Abdurrouf3} (Figure 15 therein). It is important to note that these results are sensitive to the assumption on $\alpha_{\rm CO}$. The above trends are obtained with the assumption of a constant $\alpha_{\rm CO}$. We have shown in \citetalias{2021Abdurrouf3} that the dust-to-gas mass ratio profiles deviate from flat and increase toward the galactic center when a metallicity-dependent $\alpha_{\rm CO}$ is assumed. Since the gas-phase metallicity profiles are increasing toward the galactic center in all the sample galaxies \citep{2014Pilyugin, 2020Berg}, the dust-to-gas mass ratio tends to increase with the increasing gas-phase metallicity. Given the dependency of the dust-to-gas mass ratio on the metallicity, we analyze the effect of the metallicity-dependent $\alpha_{\rm CO}$ assumption on the $\Sigma_{\rm dust}$--$\Sigma_{\rm gas}$ relation in Appendix~\ref{sec:effect_varian_alpha_CO}. Overall, we find that the scatter of this relation is slightly larger (but still considerably tight) when we assume the metallicity-dependent $\alpha_{\rm CO}$ (see Table~\ref{tab:compare_alphaCO}).

When we separate the gas into the atomic and molecular components, the dust surface density is more tightly correlated with the molecular gas compared to the atomic one. This trend agrees with the picture in which dust plays roles in catalyzing the formation of molecular gas \citep[e.g.,][]{1963Gould, 1971Hollenbach, 2004Cazaux, 2011Yamasawa}. The strong correlation between dust and $\text{H}_{2}$ has also been observed at even smaller scales ($10-50$ pc) by \citet{2014RomanDuval}, who studied the dust-to-gas mass ratio in the Magellanic Clouds. They found a higher dust-to-gas mass ratio in the dense ISM environment dominated by molecular gas. Other studies of nearby galaxies, by \citet{2012Foyle} and \citet{2014Hughes}, have also observed tight relations of $\Sigma_{\rm dust}$--$\Sigma_{\rm gas}$ and $\Sigma_{\rm dust}$--$\Sigma_{\rm H_{2}}$, but a loose relation between the $\Sigma_{\rm dust}$ and $\Sigma_{\rm HI}$, in good agreement with our results. 

On global scales, the total dust mass is also found to be correlated more strongly with the total gas mass than with either $M_{\rm HI}$ or $M_{\rm H_{2}}$ \citep[e.g.,][]{2012Corbelli, 2017Orellana, 2020Casasola}, in agreement with the trend we observe on kiloparsec scales. However, there is a disagreement among previous studies in terms of the relative significance between the $M_{\rm dust}$--$M_{\rm HI}$ and $M_{\rm dust}$--$M_{\rm H_{2}}$ relations. While \citet{2012Corbelli} and \citet{2017Orellana} found that $M_{\rm dust}$ tends to correlate more strongly with $M_{\rm H_{2}}$ than with $M_{\rm HI}$ (in agreement with the trend in sub-galactic scales that we observe), \citet{2020Casasola} found the opposite trend. Another interesting result by \citet{2018Bertemes}, who performed cross-calibration of CO- and dust-based $\text{H}_{2}$, implies that dust traces not only molecular hydrogen but also part of the atomic hydrogen in the inner molecular-dominated region, which is consistent with our spatially resolved $\Sigma_{\rm dust}$--$\Sigma_{\rm H_{2}}$--$\Sigma_{\rm HI}$ relations that mostly emerge from the regions within the $\text{H}_{2}$-dominated regime.

\begin{deluxetable}{ccccc}
\tablenum{3}
\tablecaption{Results of the Fitting of Linear Function to the Resolved Dust Scaling Relations \label{tab:fit_dust_relation}}
\tablewidth{0pt}
\tablehead{
\colhead{Correlation} & \colhead{$\rho$} & \colhead{$\alpha$} & \colhead{$\beta$} & \colhead{$\sigma$}
}
\decimals
\startdata
$\Sigma_{\rm dust}$--$\Sigma_{\rm HI}$ & $0.53$ & $0.48\pm 0.01$ & $4.17\pm 0.07$ & $0.15$ \\ 
$\Sigma_{\rm dust}$--$\Sigma_{\rm H_{2}}$ & $0.76$ & $1.48\pm 0.01$ & $-1.02\pm 0.07$ & $0.14$ \\
$\Sigma_{\rm dust}$--$\Sigma_{\rm gas}$ & $0.85$ & $0.96\pm 0.01$ & $2.06\pm 0.04$ & $0.10$ \\ 
$\Sigma_{\rm dust}$--$\Sigma_{\rm SFR}$ & $0.62$ & $2.14\pm 0.03$ & $-14.07\pm 0.19$ & $0.20$ \\ 
$\Sigma_{\rm dust}$--$\Sigma_{*}$ & $0.58$ & $2.05\pm 0.04$ & $-3.22\pm 0.21$ & $0.20$ \\ 
$\Sigma_{\rm dust}$--sSFR & $0.06$ & $28.21\pm 13.01$ & $-164.48\pm 71.08$ & $0.27$ \\ 
\enddata
%\tablecomments{}
\end{deluxetable} 

The second row in Figure~\ref{fig:edit_Mdust_MHI_MH2_bins} shows the scaling relations of $\Sigma_{\rm dust}$ with $\Sigma_{*}$, $\Sigma_{\rm SFR}$, and sSFR. While $\Sigma_{\rm dust}$ is strongly correlated with $\Sigma_{\rm SFR}$ and $\Sigma_{*}$ (as corroborated by the small $\sigma$ and high $\rho$ values), it appears to have no correlation with the sSFR on spatially resolved scales. The lack of the correlation with the sSFR may be a consequence of the linear slope ($\alpha \sim 1$) of the rSFMS relation, which suggests a broadly constant sSFR.

A previous study on the global properties of galaxies by \citet{2014Santini} suggested that the $M_{\rm dust}$--$M_{*}$ relation is likely to be a direct consequence of the SFR--$M_{\rm dust}$ and SFMS relations. They further suggest that the relationship between SFR and $M_{\rm dust}$ is a natural consequence of the KS and $M_{\rm dust}$--$M_{\rm H_{2}}$ relations. However, our results on kiloparsec scales does not necessarily support their arguments. We find that the $\Sigma_{\rm dust}$--$\Sigma_{*}$ and $\Sigma_{\rm dust}$--$\Sigma_{\rm SFR}$ relations are similarly tight (i.e., significant, as indicated by the small scatter and high $\rho$ value) and both relations are tighter than the rSFMS. Therefore, it is unlikely that the relation between the $\Sigma_{\rm dust}$ and $\Sigma_{*}$ is originated from the rSFMS. 

The color coding in Figure~\ref{fig:edit_Mdust_MHI_MH2_bins} reveals a clear radial trend in the majority of the scaling relations. The radial gradient in the $\Sigma_{\rm dust}$--$\Sigma_{\rm HI}$ appears to be perpendicular to the best-fit relation, indicating that the radial trend is mainly responsible for the scatter in this relation. This trend suggests an increasing \HI{}-to-dust mass ratio with radius (i.e., having a positive gradient). The radial trend in $\Sigma_{\rm dust}$--$\Sigma_{\rm H_{2}}$, $\Sigma_{\rm dust}$--$\Sigma_{\rm gas}$, and $\Sigma_{\rm dust}$--$\Sigma_{\rm SFR}$ show an increasing surface densities with decreasing radius, as expected from the radial profiles of these quantities (see Figure 8 in \citetalias{2021Abdurrouf3}). However, we do not see a clear radial trend in the $\Sigma_{\rm dust}$--$\Sigma_{*}$ and $\Sigma_{\rm dust}$--sSFR relations.

%%%%%%%%%%%%%%%%%%%%%%%%%%%%%%%%%%%%
\section{Variations of the Scaling Relations in Individual Galaxies} \label{sec:indiv_relations}

In order to gain a deeper understanding of the spatially resolved scaling relations discussed in the previous section, particularly regarding the origin of the scatter and the influences of the global properties of galaxies on these scaling relations, here we investigate how the scaling relations vary from galaxy to galaxy.

\subsection{Resolved Star Formation Scaling Relations in Individual Galaxies} \label{sec:res_SF_indiv}

\begin{figure*}
\centering
\includegraphics[width=0.32\textwidth]{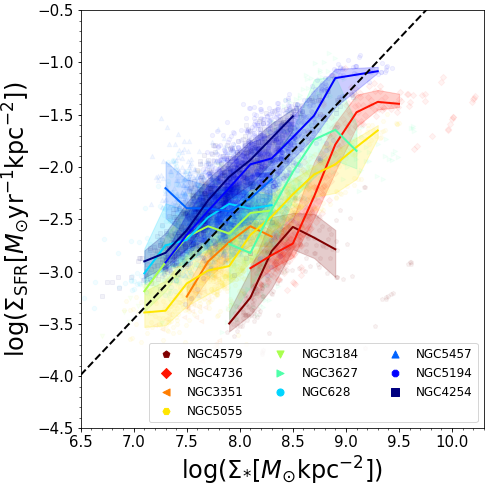}
\includegraphics[width=0.32\textwidth]{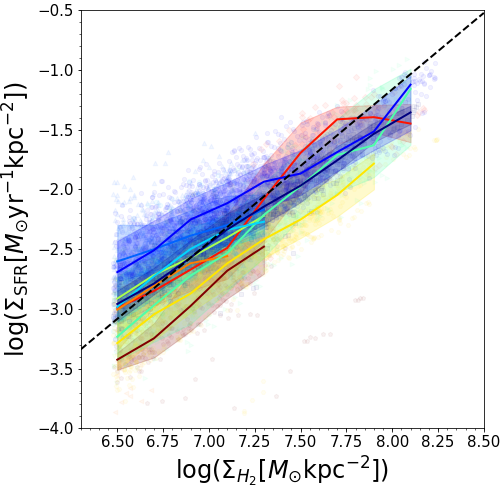}
\includegraphics[width=0.32\textwidth]{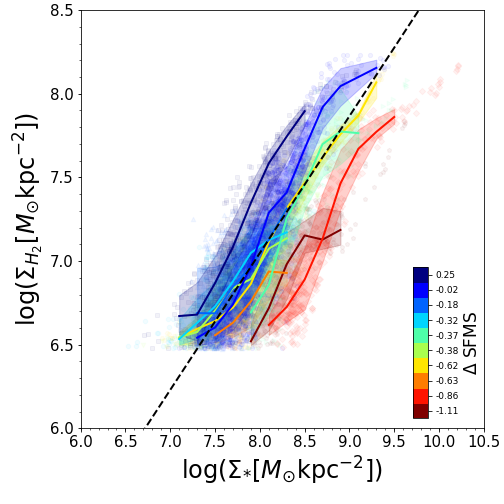}
\caption{Galaxy-to-galaxy variations of the resolved star formation scaling relations: rSFMS (left panel), rKS (middle panel), and rMGMS (right panel). Scaling relations associated with different galaxies are represented by lines and shaded areas with different colors. These individual scaling relations are derived based on the percentiles (16th, 50th, and 84th) of the distribution of the quantity in the $y$-axis for each $0.2$ dex bin of a quantity in the $x$-axis. The color-coding is based on the distance of the galaxies from the global SFMS ridge line ($\Delta$SFMS), as defined in \citetalias{2021Abdurrouf3}. The black dashed lines are the 
best-fit to the ensemble scaling relations shown in Figure~\ref{fig:merge_SF_bin_pixs}. The three relations exhibit variations from galaxy to galaxy in terms of the slope and normalization. Galaxies that have higher global SFMS normalization (i.e.,~global sSFR) tend to have higher normalization in the resolved star formation scaling relations. This indicates a significant correlations between the global sSFR and the overall level of the local sSFR, SFE, and $f_{\rm H_{2}}$ in galaxies.}
\label{fig:comb_SFMS_KS_MGMS_indiv_gals}
\end{figure*}

We show the resolved star formation scaling relations of individual galaxies in Figure~\ref{fig:comb_SFMS_KS_MGMS_indiv_gals}. In each panel, the scaling relation of a galaxy is obtained by calculating the median of the distribution of a quantity on the $y$-axis for each bin of the quantity on the $x$-axis with a bin width of $0.2$ dex. In deriving the running median, we discard the bins that contain less than 10 pixels. The scaling relations of different galaxies are shown by lines and shaded areas, color-coded by their distances from the global SFMS ridge line ($\Delta$SFMS), as defined in \citetalias{2021Abdurrouf3} (Figure 1 therein). The shaded area around a running median represents the range given by the 16th and 84th percentiles. The black dashed lines are the best-fit to the ensemble scaling relations as shown in Figure~\ref{fig:merge_SF_bin_pixs}. 

All the three relations exhibit variations from galaxy to galaxy in terms of the normalization, slope, and shape. The rSFMS relation shows the most significant variations among the three, while the rKS relation has the least variations. The small variation in the slope of the rKS relation may imply a universal star formation law, while the variation in the normalization of this relation reflects the diversity in the local SFE among the galaxies.

The scatter of the relations in individual galaxies is significantly smaller than the scatter of the ensemble relation, indicating that the scatter in the ensemble relation is mainly driven by the galaxy-to-galaxy variation (i.e.,~differences in slope and normalization). Considering the radial profiles of the $\Sigma_{*}$, $\Sigma_{\rm SFR}$, and $\Sigma_{\rm H_{2}}$ \citepalias[][Figure~8 therein]{2021Abdurrouf3}, which decrease with increasing galactocentric distances, the pixels from the bottom left to the top right sides of each relation (of individual galaxies) tend to have decreasing radii.

As can be seen from the left panel of Figure~\ref{fig:comb_SFMS_KS_MGMS_indiv_gals}, the rSFMS relation shows a saturation of $\Sigma_{\rm SFR}$ at high $\Sigma_{*}$ for most of the galaxies. Such a flattening has also been observed by previous studies \citep[e.g.,][]{2017Abdurrouf, 2018Abdurrouf, 2021Ellison}, and is mainly caused by the central quiescent bulge component. A similar flattening trend is also found in the rMGMS relation for most of the galaxies. This flattening trend is likely a consequence of the suppression of the molecular gas-to-stellar mass ratio in the central regions of the galaxies, as we observed in \citetalias{2021Abdurrouf3} (Figure~12 therein). A previous study by \citet{2021Ellison} also observed a similar saturation of $\Sigma_{\rm H_{2}}$ in high $\Sigma_{*}$ regions of the rMGMS relation.

The color-coding clearly shows that galaxies with higher global SFMS normalization (i.e.,~higher global sSFR) tend to have higher normalization in the three resolved star formation scaling relations. This trend suggests that the normalization of these scaling relations, which represents the overall levels of the sSFR, SFE, and $\text{H}_{2}$ fraction on local kiloparsec scales within galaxies, is likely correlated with some global processes that determine the global sSFR. Based on these results, we can also see that the diversity in the global sSFR contributes to the scatter of the resolved star formation scaling relations formed with the ensemble of all pixels  (see Section~\ref{sec:merged_SF_relations}).

%%%%%%%%%%%%%%%%%%%%%%%%%%
\subsection{Resolved Dust Scaling Relations in Individual Galaxies}

\begin{figure*}
\centering
\includegraphics[width=0.33\textwidth]{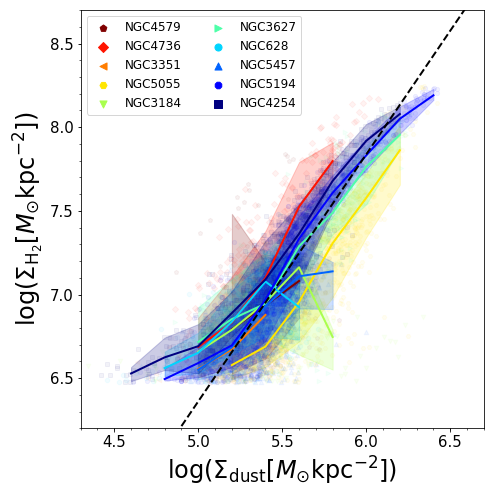}
\includegraphics[width=0.33\textwidth]{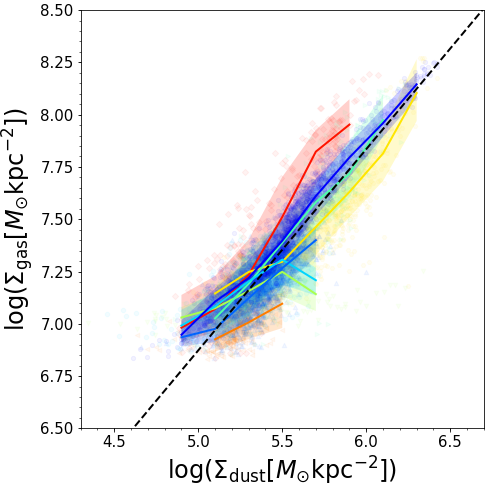}
\includegraphics[width=0.33\textwidth]{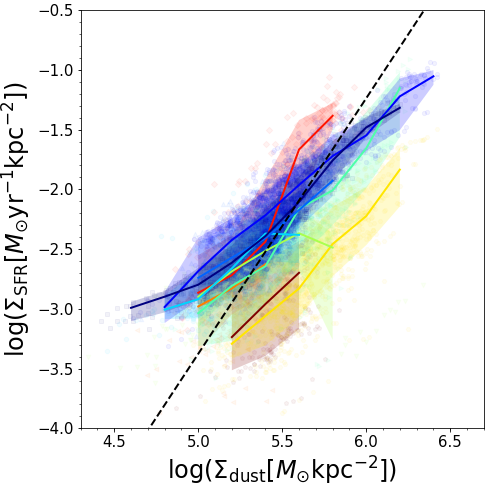}
\includegraphics[width=0.33\textwidth]{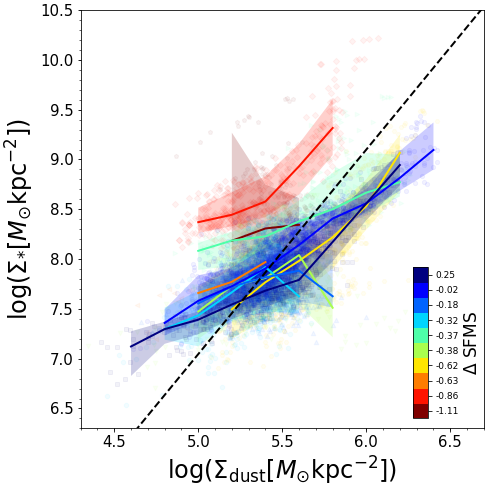}
\caption{Galaxy-to-galaxy variations of the resolved dust scaling relations. The symbols in this figure is the same as those in Figure~\ref{fig:comb_SFMS_KS_MGMS_indiv_gals}. Two galaxies (NGC 4254 and NGC 4579) are not included in the $\Sigma_{\rm dust}$--$\Sigma_{\rm gas}$ relation because they do not have \HI{} data from the THINGS survey. The $\Sigma_{\rm dust}$--$\Sigma_{\rm H_{2}}$ and $\Sigma_{\rm dust}$--$\Sigma_{\rm gas}$ relations exhibit very little variation from galaxy to galaxy, which indicates that these relations are likely universal for galaxies with a wide range of global sSFRs. A more significant variations are observed in the $\Sigma_{\rm dust}$--$\Sigma_{\rm SFR}$ and $\Sigma_{\rm dust}$--$\Sigma_{*}$ relations, where we see a systematic effect of the global sSFR on the normalization of these resolved relations.}
\label{fig:edit_Mdust_MHI_MH2_SM_pix_indiv}
\end{figure*}

Next, we investigate the galaxy-to-galaxy variations of the resolved dust scaling relations. The dust scaling relations of individual galaxies are shown in Figure~\ref{fig:edit_Mdust_MHI_MH2_SM_pix_indiv}. We only analyze four relations ($\Sigma_{\rm dust}$--$\Sigma_{\rm H_{2}}$, $\Sigma_{\rm dust}$--$\Sigma_{\rm gas}$, $\Sigma_{\rm dust}$--$\Sigma_{\rm SFR}$, and $\Sigma_{\rm dust}$--$\Sigma_{*}$) as the other two relations ($\Sigma_{\rm dust}$--$\Sigma_{\rm HI}$ and $\Sigma_{\rm dust}$--sSFR) are weak (see Figure~\ref{fig:edit_Mdust_MHI_MH2_bins}). As can be seen from the figure, the $\Sigma_{\rm dust}$--$\Sigma_{\rm H_{2}}$ and $\Sigma_{\rm dust}$--$\Sigma_{\rm gas}$ relations of individual galaxies are located in the same tight locus, which indicates that this relation holds similarly among our sample galaxies. We note that our sample is limited to relatively massive spiral galaxies and thus may not be representative of the general population of galaxies. Although our sample covers a wide range of global sSFR, it does not cover a sufficiently wide range of gas-phase metallicity. Referring to \citet{2010Moustakas}, our sample galaxies have a global $12+\log(\text{O}/\text{H})$ that ranges from $8.31$ to $8.68$, if the \citet{2005Pilyugin} calibration is adopted, or from $8.99$ to $9.22$, if the calibration of \citet{2004Kobulnicky} is applied. Our sample only covers the high-mass plateau in the mass-metallicity relation of the SINGS galaxies as analyzed in \citet[][Figure 11 therein]{2010Moustakas}. One may expect that the $\Sigma_{\rm dust}$--$\Sigma_{\rm gas}$ relation to be influenced by the gas-phase metallicity because of the known relationship between the dust-to-gas mass ratio and the gas-phase metallicity \citep[e.g.,][]{2014RemyRuyer,2018Chiang}. The influence of metallicity could come from the dependency of $\alpha_{\rm CO}$ on the metallicity. To check this, in Appendix~\ref{sec:effect_varian_alpha_CO}, we recalculate $\Sigma_{\rm H_{2}}$ of the sub-galactic regions with a metallicity-dependent $\alpha_{\rm CO}$, and find that the above two scaling relations still hold with slightly larger scatter and shallower slope. 

In the second row of Figure~\ref{fig:edit_Mdust_MHI_MH2_SM_pix_indiv}, we see that the $\Sigma_{\rm dust}$--$\Sigma_{\rm SFR}$ and $\Sigma_{\rm dust}$--$\Sigma_{*}$ relations exhibit a more significant variation in normalization than the other two relations. Similar to the case of the resolved star formation scaling relations (Section~\ref{sec:res_SF_indiv}), these resolved dust scaling relations in individual galaxies are tighter than the corresponding ensemble scaling relations (see Section~\ref{sec:merged_dust_relations}) and it is the galaxy-to-galaxy variations in normalization of these relations that mainly contributes to the scatter of the ensemble scaling relations. As can be seen from the color-coding, galaxies that are more actively forming stars tend to have higher  (lower) normalization in the $\Sigma_{\rm dust}$--$\Sigma_{\rm SFR}$  ($\Sigma_{\rm dust}$--$\Sigma_{*}$) relations.

\section{Correlations Among the Scaling Relations}
\label{sec:correlation_scaling0}

In Section~\ref{sec:indiv_relations}, we have shown that the spatially resolved star formation scaling relations exhibit a significant galaxy-to-galaxy variations, which is most prominently seen in the normalization of the relations. In this section, we compare the normalization of those scaling relations in individual galaxies to get further insight on the correlations among the scaling relations. First, we compare the normalization of the three resolved star formation scaling relations in Section~\ref{sec:correlation_scaling}, then we investigate correlations between the normalization of the star formation and dust scaling relations in Section~\ref{sec:correlation_dust_SF_scaling}.

\subsection{Correlations Among the Resolved Star Formation Scaling Relations}
\label{sec:correlation_scaling}

\begin{figure}
\centering
\includegraphics[width=0.4\textwidth]{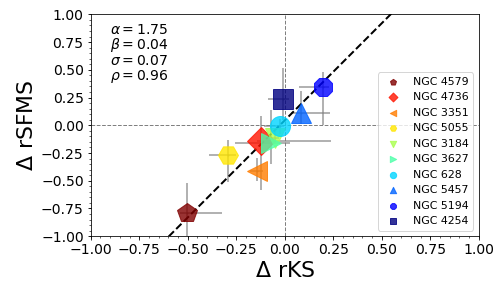}
\includegraphics[width=0.4\textwidth]{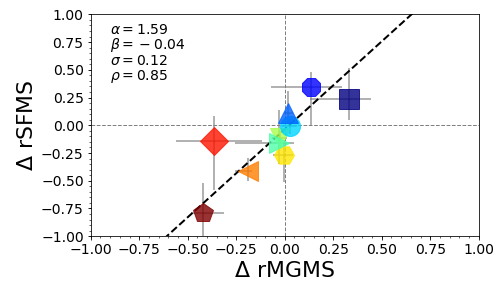}
\includegraphics[width=0.4\textwidth]{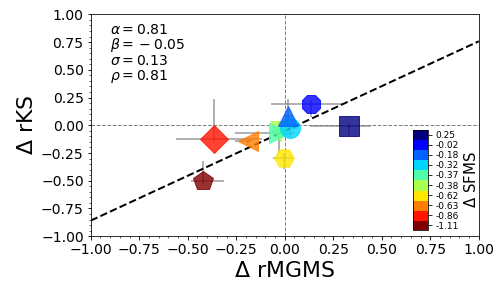}
\caption{Correlations among the normalization offsets of the resolved star formation scaling relations of individual galaxies with respect to the ensemble relations formed with all spatial regions ($\Delta$rSFMS, $\Delta$rKS, and $\Delta$rMGMS). For a galaxy, the offset is defined as the median of the offsets between the median relation associated with the galaxy and that of the ensemble relation. The color-coding represents the distance from the global SFMS ($\Delta$SFMS). There are significant correlations among $\Delta$rSFMS, $\Delta$rKS, and $\Delta$rMGMS, which overall suggest that galaxies with higher level of resolved $f_{\rm H_{2}}$ tend to have higher levels of resolved SFE and sSFR.}
\label{fig:dSFMS_dKS_dMGMS_comp}
\end{figure}

We start by quantifying the offset of each scaling relation in individual galaxies with respect to the median profile of the corresponding ensemble scaling relation that is constructed from all sub-galactic regions of the sample galaxies (see Section~\ref{sec:ensemble_relations}). To do this, we calculate the distance along the $y$-axis between the median profile of the scaling relation in each galaxy and that of the ensemble relation. Then we define the offset of each relation as the median of these distances. The uncertainty of the offset is defined as the range given by the 16th and 84th percentiles. We compare the offsets of the resolved star formation scaling relations ($\Delta$rSFMS, $\Delta$rKS, and $\Delta$rMGMS) in Figure~\ref{fig:dSFMS_dKS_dMGMS_comp}. In each panel, different galaxies are indicated with different symbols and colors that represent distance from the global SFMS (as we have done in Section~\ref{sec:indiv_relations}).

We find significant positive correlations among the normalization offsets of the scaling relations: $\Delta$rKS--$\Delta$rSFMS, $\Delta$rMGMS--$\Delta$rSFMS, and $\Delta$rMGMS--$\Delta$rKS. $\Delta$rSFMS is strongly correlated with $\Delta$rKS with a high $\rho$ value of $0.96$ and a small scatter of $0.07$ dex, indicating that galaxies with higher rKS normalization tend to have higher normalization of the rSFMS relation. This means that galaxies with a higher level of resolved SFE tend to have a higher level of resolved sSFR and vice versa. $\Delta$rSFMS is also strongly correlated with $\Delta$rMGMS, although it is less significant compared to the $\Delta$rKS--$\Delta$rSFMS correlation. The $\sigma$ and $\rho$ values of the $\Delta$rMGMS--$\Delta$rSFMS relation are $0.12$ dex and $0.85$, respectively. This correlation indicates that the rate of stellar mass growth (by star formation) is also strongly influenced by the molecular gas mass fraction, such that a higher molecular gas fraction leads to a higher rate of stellar mass growth. These results are reminiscent of trends seen globally \citep[][and reference therein]{2022Saintonge}.

In the bottom panel of Figure~\ref{fig:dSFMS_dKS_dMGMS_comp}, we see that $\Delta$rMGMS and $\Delta$rKS are positively correlated, although the significance level is not as high as the other two correlations, as indicated by the larger scatter ($\sigma=0.13$ dex) and lower $\rho$ value ($0.81$) in comparison to the other two correlations. This indicates that the SFE is positively correlated with the molecular gas mass fraction, such that galaxies with more molecular gas mass supplies tend to have a higher SFE. A similar comparison of the normalization offsets of the rSFMS, rKS, and rMGMS relations by \citet{2021Ellison} found strong positive correlations in the $\Delta$rKS--$\Delta$rSFMS and $\Delta$MGMS--$\Delta$rSFMS relations, which agree with our result. However, they found a very weak correlation between the $\Delta$rMGMS and $\Delta$rKS, in contradiction with our result. This disagreement can be caused by several factors, including the sample selection and methods for deriving SFRs and stellar masses. Nevertheless, it is worth noting that we also find the weakest correlation in the $\Delta$rMGMS--$\Delta$rKS correlation than in the other two.

\begin{figure*}
\centering
\includegraphics[width=0.32\textwidth]{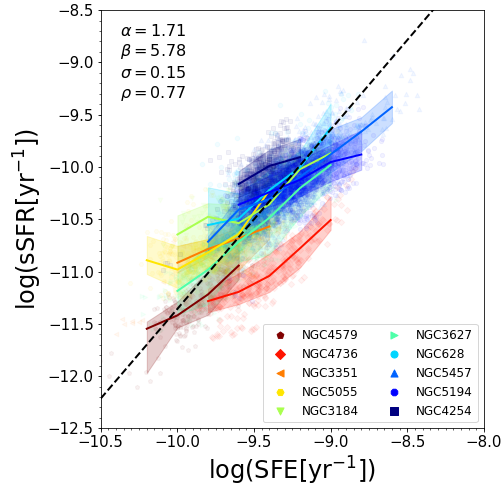}
\includegraphics[width=0.32\textwidth]{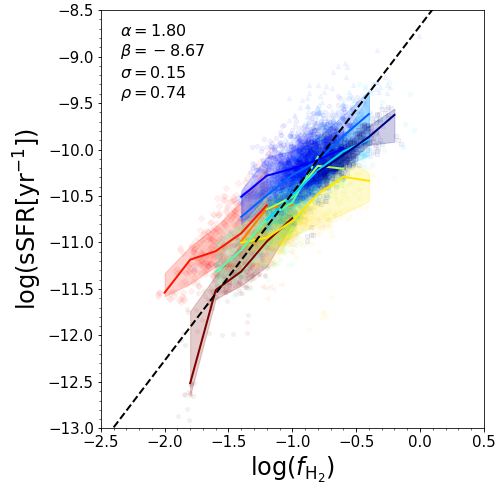}
\includegraphics[width=0.32\textwidth]{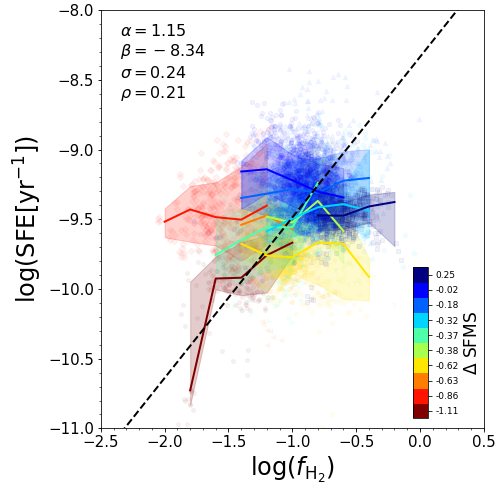}
\caption{Spatially resolved correlations among sSFR, SFE, and $f_{\rm H_{2}}$. Sub-galactic regions associated with galaxies of different $\Delta$SFMS (i.e.,~global sSFR) are tend to be separated diagonally, which indicates that the global sSFR, $f_{\rm H_{2}}$, and SFE are preserved on kiloparsec scales. In contrast to the strong positive $\Delta$rMGMS--$\Delta$rKS correlation, we see no correlation between spatially resolved $f_{\rm H_{2}}$ and SFE in individual galaxies. This may indicate that the SFE is regulated by some processes that act on global scales. The symbols in this figure is the same as those in Figure~\ref{fig:comb_SFMS_KS_MGMS_indiv_gals}. The best-fit linear function parameters are shown in the top left corner of each panel.}
\label{fig:fH2_SFE_sSFR_indiv}
\end{figure*}

Basically, we can think of $\Delta$rSFMS, $\Delta$rKS, and $\Delta$rMGMS of individual galaxies as measures of the overall levels of the spatially resolved sSFR, SFE, and $\text{H}_{2}$ mass fraction ($f_{\rm H_{2}}$) in  galaxies, respectively. The latter three quantities are related via
\begin{equation}
\text{sSFR} = \text{SFE}\times f_{\rm H_{2}}.
\label{eq:sSFR_SFE_fH2}
\end{equation}
Although $\Delta$rSFMS, $\Delta$rKS, and $\Delta$rMGMS tell about the normalization of the resolved scaling relations, they are actually global-scale quantities. To investigate the correlations among equivalent quantities of those normalization on kiloparsec scales, we plot the spatially resolved sSFR--SFE--$f_{\rm H_{2}}$ correlations in Figure~\ref{fig:fH2_SFE_sSFR_indiv}. In the left and middle panels, we see that there is a tendency of the sub-galactic regions of galaxies with different $\Delta$SFMS (i.e.,~global sSFR) to be separated diagonally, in contrast to the resolved star formation and dust scaling relations, where the regions associated with the star-forming and quiescent galaxies are tend to be separated vertically. This trend indicates that the global sSFR, SFE, and $f_{\rm H_{2}}$ are preserved on kiloparsec scales. However, there seems to be no correlation between SFE and $f_{\rm H_{2}}$ in individual galaxies, with SFE stays broadly constant over an order of magnitude change in the $\text{H}_{2}$ mass fraction. This is in contrast to the trend we see in the equivalent relation on global scales ($\Delta$rMGMS--$\Delta$rKS), where we see a significant positive correlation. This trend may indicate that the SFE is roughly constant in the disk and it is likely governed by some processes that act on global scales.

%%%%%%%%%%%%%%%%%%%%%%%%%%%%%%
\subsection{Correlation Between the Dust and Star Formation Scaling Relations}
\label{sec:correlation_dust_SF_scaling}

\begin{figure*}
\centering
\includegraphics[width=0.32\textwidth]{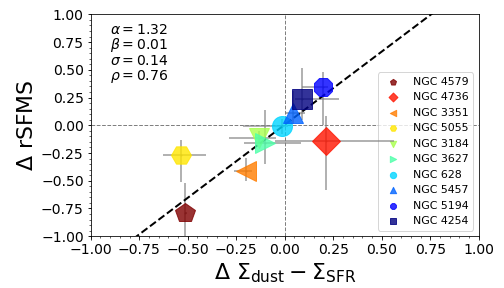}
\includegraphics[width=0.32\textwidth]{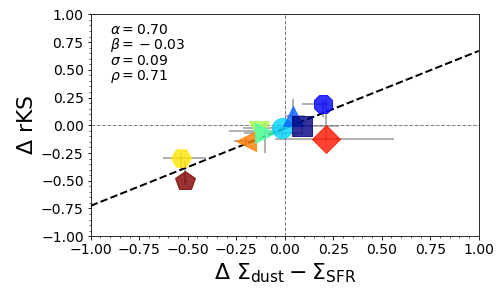}
\includegraphics[width=0.32\textwidth]{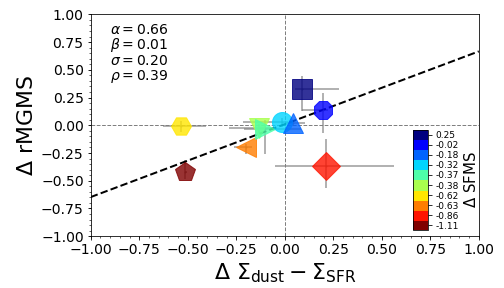}
\includegraphics[width=0.32\textwidth]{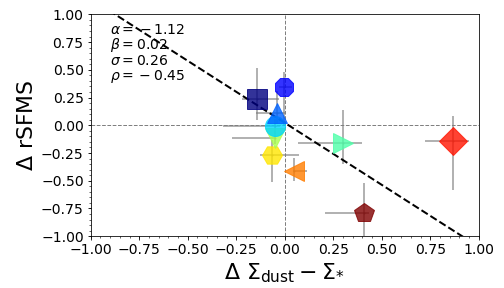}
\includegraphics[width=0.32\textwidth]{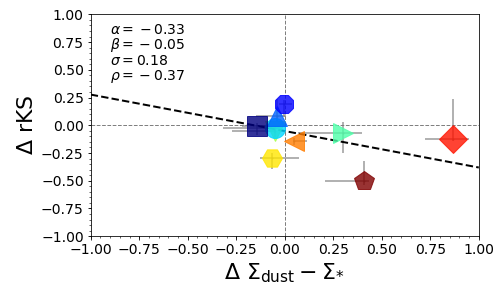}
\includegraphics[width=0.32\textwidth]{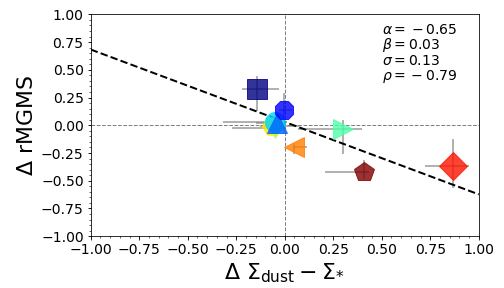}
\caption{Comparisons between the normalization offsets of the spatially resolved star formation and the dust scaling relations. Among the dust scaling relations, we only consider the $\Sigma_{\rm dust}$--$\Sigma_{\rm SFR}$ (first row) and $\Sigma_{\rm dust}$--$\Sigma_{*}$ (second row) relations because these two are significant yet exhibit galaxy-to-galaxy variations in terms of normalization. The symbols and color-coding in this figure are the same as those in Figure~\ref{fig:dSFMS_dKS_dMGMS_comp}. We observe an indication that galaxies with higher average SFR-to-dust mass ratio over their spatial regions tend to have higher averages sSFR and SFE and vice versa. There is also an indication of decreasing $f_{\rm H_{2}}$ and sSFR with decreasing dust-to-stellar mass ratio.}
\label{fig:dSF_ddust_comp}
\end{figure*}

Next, we investigate possible correlations between the resolved star formation and dust scaling relations, to examine how dust is associated with star formation activities on kiloparsec scales. For the dust scaling relations, we only consider the $\Sigma_{\rm dust}$--$\Sigma_{*}$ and $\Sigma_{\rm dust}$--$\Sigma_{\rm SFR}$ relations because they exhibit stronger galaxy-to-galaxy variations than the $\Sigma_{\rm dust}$--$\Sigma_{\rm H_{2}}$ and $\Sigma_{\rm dust}$--$\Sigma_{\rm gas}$ relations, while the other two relations ($\Sigma_{\rm dust}$--$\Sigma_{\rm HI}$ and $\Sigma_{\rm dust}$--sSFR) are weak (see Figure~\ref{fig:edit_Mdust_MHI_MH2_bins}). In Figure~\ref{fig:dSF_ddust_comp}, we show correlations of the normalization offsets between the resolved star formation scaling relations and the above two dust scaling relations. It can be seen that the normalization offset of the $\Sigma_{\rm dust}$--$\Sigma_{\rm SFR}$ relation is strongly correlated ($\sigma<0.15$ dex and $\rho>0.7$) with the offsets of the rSFMS and rKS relations, but only weakly correlated with the offset of the rMGMS relation ($\rho=0.39$). On the other hand, the normalization offset of the $\Sigma_{\rm dust}$--$\Sigma_{*}$ relation has a strong negative correlation with the offset of the rMGMS relation, but has a relatively weak negative correlation with the offsets of the rSFMS and rKS relations. 

The $\Delta \Sigma_{\rm dust}$--$\Sigma_{\rm SFR}$ vs. $\Delta$rSFMS and $\Delta \Sigma_{\rm dust}$--$\Sigma_{\rm SFR}$ vs. $\Delta$rKS correlations indicate that the galaxies that have higher average SFR-to-dust mass ratio over their spatial regions tend to have higher averages sSFR and SFE. The $\Delta \Sigma_{\rm dust}$--$\Sigma_{*}$ vs. $\Delta$rMGMS relation suggests that the galaxies that have lower average dust-to-stellar mass ratio tend to have lower average $f_{\rm H_{2}}$. There is also a hint of decreasing sSFR with decreasing dust-to-stellar mass ratio, as indicated by the $\Delta \Sigma_{\rm dust}$--$\Sigma_{*}$ vs. $\Delta$rSFMS relation. Looking at the quantities involved in the above relations, one may argue that the strong relationship is just the consequence of the fact that the two axes share a same quantity. However, this is not always the case. As we have seen in the right panel of Figure~\ref{fig:fH2_SFE_sSFR_indiv}, although both axes in the $f_{\rm H_{2}}$--SFE plane share the same quantity $\Sigma_{\rm H_{2}}$, there is no clear correlation on this plane.

\begin{figure}
\centering
\includegraphics[width=0.4\textwidth]{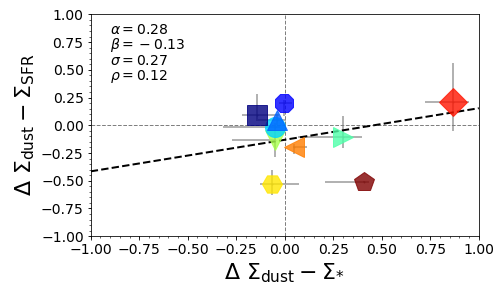}
\caption{Comparison between the normalization offsets of the $\Sigma_{\rm dust}$--$\Sigma_{*}$ and $\Sigma_{\rm dust}$--$\Sigma_{\rm SFR}$ relations in individual galaxies. The symbols and color-coding in this figure are the same as those in Figure~\ref{fig:dSFMS_dKS_dMGMS_comp}. There seems to be no correlation between the normalization offsets of the $\Sigma_{\rm dust}$--$\Sigma_{*}$ and $\Sigma_{\rm dust}$--$\Sigma_{\rm SFR}$ relations. This indicates that these two relations are independent of each other.}
\label{fig:dSMMdust_vs_dMdustSFR}
\end{figure}

After having examined the connections between the resolved dust and star formation scaling relations, now we investigate the possible connection between $\Sigma_{\rm dust}$--$\Sigma_{*}$ and $\Sigma_{\rm dust}$--$\Sigma_{\rm SFR}$ relations. We show the comparison between the normalization offsets of the $\Sigma_{\rm dust}$--$\Sigma_{*}$ and $\Sigma_{\rm dust}$--$\Sigma_{\rm SFR}$ in Figure~\ref{fig:dSMMdust_vs_dMdustSFR}. There appears to be no correlation between $\Delta \Sigma_{\rm dust}$--$\Sigma_{*}$ and $\Delta \Sigma_{\rm dust}$--$\Sigma_{\rm SFR}$, suggesting that these two relations tend to be independent of each other, in the sense that the $\Sigma_{\rm dust}$--$\Sigma_{*}$ relation is not the consequence of the $\Sigma_{\rm dust}$--$\Sigma_{\rm SFR}$ and rSFMS relations, nor is the $\Sigma_{\rm dust}$--$\Sigma_{\rm SFR}$ relation the consequence of the $\Sigma_{\rm dust}$--$\Sigma_{\rm H_{2}}$ and rKS relations. This weak correlation between the $\Delta \Sigma_{\rm dust}$--$\Sigma_{*}$ and $\Delta \Sigma_{\rm dust}$--$\Sigma_{\rm SFR}$ also supports our earlier argument that involving a same quantity into both axes does not always results in a strong relationship.

\section{Discussion}
\label{sec:discussion}

\subsection{What Drives the Galaxy-to-galaxy Variation of the Resolved Star Formation Scaling Relations?} \label{sec:origin_scatter}

\begin{figure*}
\centering
\includegraphics[width=0.32\textwidth]{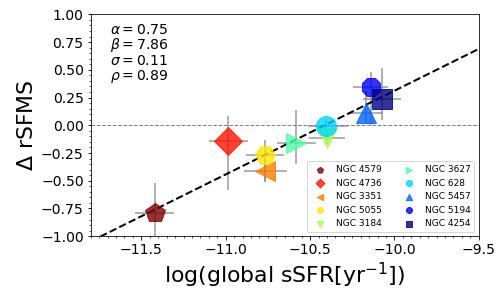}
\includegraphics[width=0.32\textwidth]{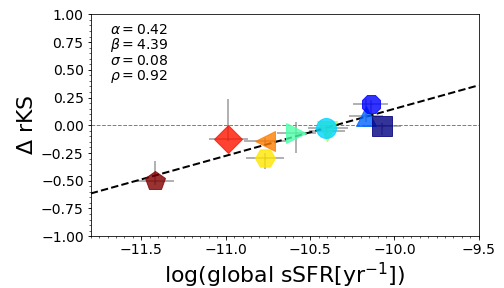}
\includegraphics[width=0.32\textwidth]{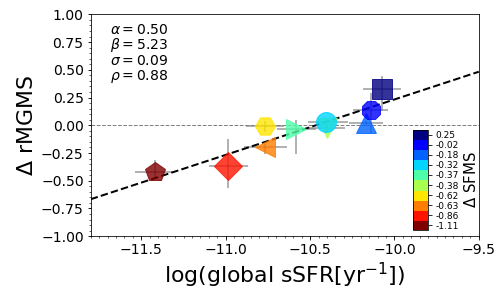}
\includegraphics[width=0.32\textwidth]{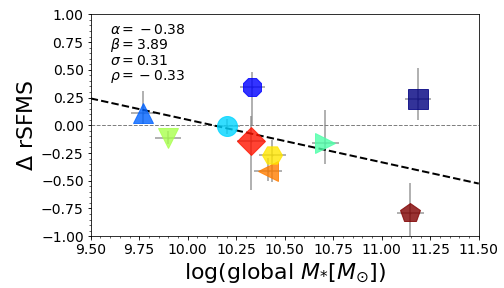}
\includegraphics[width=0.32\textwidth]{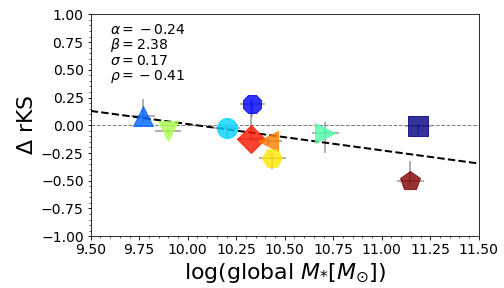}
\includegraphics[width=0.32\textwidth]{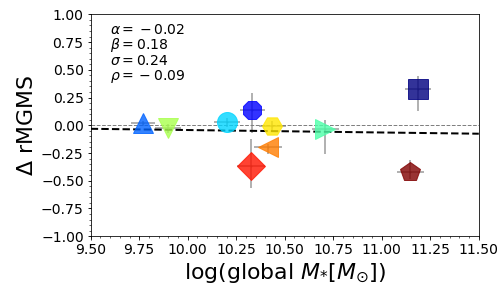}
\includegraphics[width=0.32\textwidth]{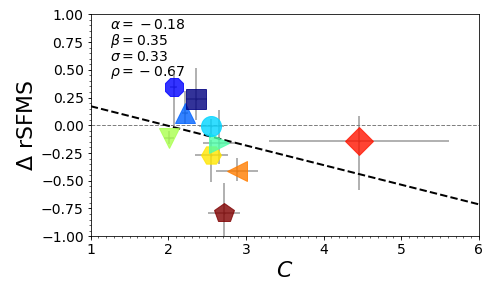}
\includegraphics[width=0.32\textwidth]{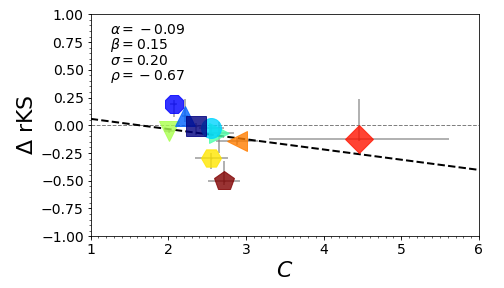}
\includegraphics[width=0.32\textwidth]{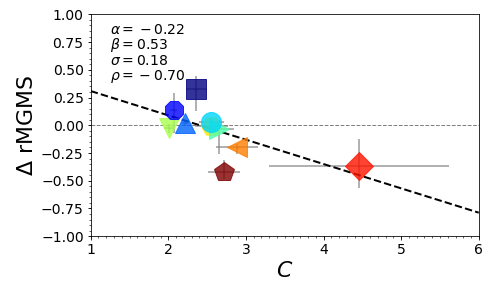}
\caption{Effects of the global sSFR (first row), $M_{*}$ (second row), and concentration index ($C$; third row) on the normalization offsets of the resolved star formation scaling relations of the individual galaxies with respect to the median profiles of the ensemble star formation scaling relations ($\Delta$rSFMS, $\Delta$rKS, and $\Delta$rMGMS). The concentration index is defined as $C\equiv R_{90}/R_{50}$, where $R_{90}$ and $R_{50}$ are the radii that enclose $90\%$ and $50\%$ of the total stellar mass. The symbols and color-coding are the same as those in Figure~\ref{fig:dSFMS_dKS_dMGMS_comp}. The normalization offsets of the rSFMS, rKS, and rMGMS relations are strongly correlated with the global sSFR, while they seem to be not correlated with the global $M_{*}$. This trend indicates that galaxies with higher global sSFR tend to have higher level of local sSFR, SFE, and $f_{\rm H_{2}}$ over their spatial regions. We also observe that the normalization offsets are correlated with the concentration index in such a way that more compact galaxies tend to have lower level of local sSFR, SFE, and $f_{\rm H_{2}}$ over their spatial regions.}
\label{fig:edit_SMs_sSFR_delta_scaling}
\end{figure*}

As shown in Section~\ref{sec:res_SF_indiv}, the resolved star formation scaling relations exhibit variations from galaxy to galaxy (Figure~\ref{fig:comb_SFMS_KS_MGMS_indiv_gals}). To investigate the possible causes for the galaxy-to-galaxy variations, we check correlations between the normalization of the rSFMS, rKS, and rMGMS relations and the global properties of galaxies. Figure~\ref{fig:edit_SMs_sSFR_delta_scaling} shows the normalization offsets of the sample galaxies in the rSFMS, rKS, and rMGMS relations as functions of the global sSFR (first row), $M_{*}$ (second row), and concentration index ($C$; third row). The concentration index, which measures the compactness of the stellar distribution in a galaxy, is defined as $C\equiv R_{90}/R_{50}$, where $R_{90}$ and $R_{50}$ are the galactocentric distances (measured along the elliptical semi-major axis) that enclose $90\%$ and $50\%$ of the total stellar mass, respectively. Generally, galaxies with early-type morphology tend to have higher $C$ than those with late-type morphology \citep[e.g.,][]{2001Shimasaku, 2001Strateva, 2003Nakamura, 2005Park}. Previous studies have also shown that $C$ correlates positively with bulge-to-total ratio \citep[$B/T$; e.g.,][]{2016Kim}. 

From the first row in Figure~\ref{fig:edit_SMs_sSFR_delta_scaling} we observe that the normalization offsets of the three resolved star formation scaling relations are strongly correlated with the global sSFR in such a way that galaxies with higher sSFR tend to have higher normalization of rSFMS, rKS, and rMGMS relations. Among these three, the sSFR is correlated most strongly with $\Delta$rKS ($\sigma=0.08$ dex and $\rho=0.92$), suggesting a strong positive correlation between the global sSFR and the SFE within galaxies. One may expect the sSFR to be more strongly correlated with $\Delta$rSFMS than with $\Delta$rKS because both of these quantities are different expressions of the same physical property. However, our result shows that this is not necessarily the case. The strong positive sSFR--$\Delta$rMGMS correlation suggests that there is a tight relation between the global sSFR and the overall level of the molecular gas fraction in galaxies. The color-coding shows a clear trend of an increasing normalization of the global SFMS relation from the bottom left side to the top right side on these three diagrams. A similar analysis carried out by \citet{2021Ellison}, using a larger sample of low-$z$ galaxies, also obtained a strong positive correlations between the global sSFR and $\Delta$rSFMS, $\Delta$rKS, and $\Delta$rMGMS, in good agreement with our results.      

In the second row of Figure~\ref{fig:edit_SMs_sSFR_delta_scaling}, we observe weak correlations between the global $M_{*}$ and the normalization offsets of the resolved star formation scaling relations. While $\Delta$rMGMS seems to be constant across a wide range of $M_{*}$ (see the right panel), there is a tendency of decreasing normalization in the rSFMS and rKS relations with increasing $M_{*}$, as shown in the left and middle panels. These results are again in good agreement with the finding of \citet{2021Ellison}.

The third row of Figure~\ref{fig:edit_SMs_sSFR_delta_scaling} shows correlations between the concentration index and the normalization offsets. Despite the small size and the limited morphological type (i.e.~only spirals) of our sample, we still observe clear correlations between $C$ and the normalization offsets of the scaling relations. The correlations are all significant, with $|\rho| \gtrsim 0.67$. This result suggests that more compact galaxies (i.e.,~having higher $C$) tend to have lower levels of resolved sSFR, SFE, and molecular gas fraction. The suppressed $\text{H}_{2}$ fraction and SFE in galaxies with higher $C$ (i.e.~higher $B/T$) seem to support the scenario in which the star formation suppression is promoted by the existence of the massive bulge component \citep{2009Martig}.

Previous studies have also explored correlations between galaxy morphology and star formation scaling relations on both global and kiloparsec scales. On global scales, early-type galaxies tend to reside on the lower envelope or below the SFMS, while late-type galaxies are mostly located on the upper envelope and the SFMS itself \citep[e.g.,][]{2011Wuyts, 2016GonzalezDelgado, 2019CanoDiaz}. For the global $M_{\rm H_{2}}$--SFR relation, the $B/T$ ratio tends to increase with decreasing SFR at a given $M_{\rm H_{2}}$ \citep{2021Duo}. A correlation between $\text{H}_{2}$ depletion time and $C$ that is observed by \citet{2012Saintonge} is also consistent with the picture that more compact galaxies tend to have longer $\text{H}_{2}$ depletion time (i.e.~lower SFE). Because the MGMS relation has been studied in detail only recently, its correlations with the global properties is not well understood. However, we can regard the MGMS normalization as the $\text{H}_{2}$ fraction, which is better studied. In line with this, \citet{2011Saintonge} observed a rather weak negative correlation between the $C$ and the $\text{H}_{2}$ fraction. On kiloparsec scales, \citet{2021Ellison} found strong negative correlations between the Sersic index and the normalization offsets of the rSFMS and rKS relations, in agreement with our results. However, they found no correlation between the Sersic index and $\Delta$rMGMS.

%%%%%%%%%%%%%%%%%%%%%%%%%%%%%%%%%%%%
\subsection{What are the Conditions for the \HI{}-to-H$_{2}$ Transition?}
\label{sec:driver_mol_formation}

Based on our current understanding of the galaxy formation and evolution, the star formation in galaxies is regulated by gas flow in and out of the galaxies \citep[e.g.,][]{2006Dekel, 2010Bouche, 2010Tacconi, 2013Lilly, 2016Tacchella}. The accreted gas from the cosmic web cools down to form atomic hydrogen first, and then condense to form molecular hydrogen on dust surfaces \citep[e.g.,][]{1963Gould, 1971Hollenbach, 2004Cazaux}. Star formation can occur when the molecular gas collapses owing to gravitational instability \citep{1989Kennicutt, 1998Kennicutt_b, 2001Martin}. It has been known that $\text{H}_{2}$ is centrally concentrated, while \HI{} is spatially extended (i.e.~more abundance in the outskirt regions) in galaxies (e.g.,~\citealt{1991Young}, \citealt{2008Bigiel}, \citealt{2008Leroy}). As we have shown in \citetalias{2021Abdurrouf3}, the radius of transition from the \HI{}-dominated to $\text{H}_{2}$-dominated regions varies from galaxy to galaxy. Here, we examine whether this transition occurs in a well-defined local condition, especially in terms of the local surface densities of dust, gas, and stars.

\begin{figure*}
\centering
\includegraphics[width=0.32\textwidth]{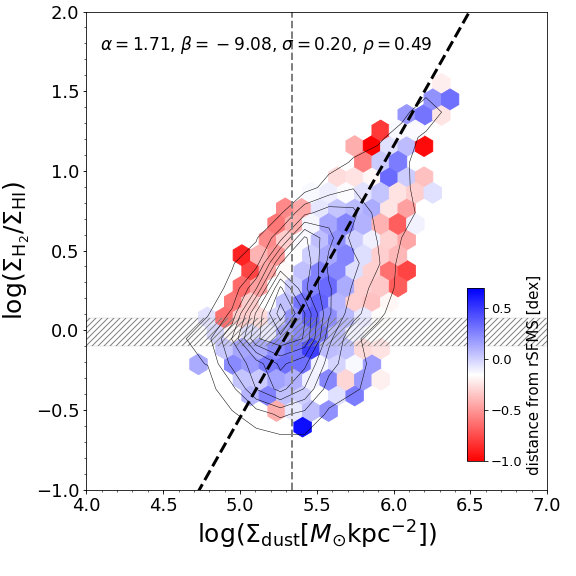}
\includegraphics[width=0.32\textwidth]{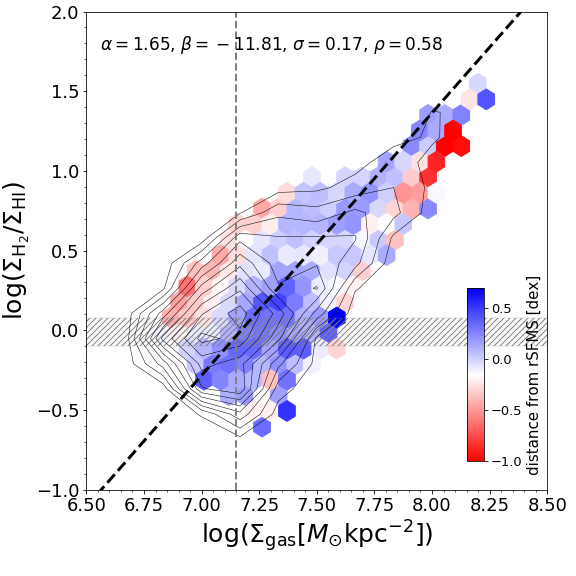}
\includegraphics[width=0.32\textwidth]{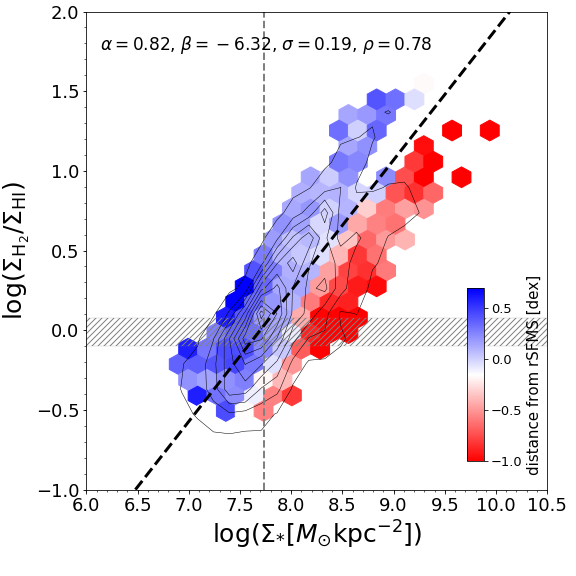}
\caption{The $\text{H}_{2}$-to-\HI{} ratio as functions of the surface densities of dust (left panel), gas (middle), and stellar mass (right panel). Only hexagonal bins that have at least three pixels are shown here. Each bin is color coded based on the average perpendicular distance from the best-fit rSMFS of the member pixels. The black dashed lines represent the best-fit of the relations, while the density contours represent the distributions of the pixels on the diagrams. The vertical dashed lines show a critical density at the \HI{}-to-$\text{H}_{2}$ transition, estimated from the median of surface densities within $0.8<\Sigma_{\rm H_{2}}/\Sigma_{\rm HI}<1.2$ (the horizontal bands). The typical ISM surface densities at the \HI{}-to-$\text{H}_{2}$ transition are: $\log(\Sigma_{\rm dust} [M_{\odot}\text{ kpc}^{-2}])=5.34\pm 0.02$, $\log(\Sigma_{\rm gas} [M_{\odot}\text{ kpc}^{-2}])= 7.15\pm 0.02$, and $\log(\Sigma_{*} [M_{\odot}\text{ kpc}^{-2}])= 7.73\pm 0.02$.}
\label{fig:mol_to_neutral_ratio}
\end{figure*}

Figure~\ref{fig:mol_to_neutral_ratio} shows the relationship between the molecular-to-neutral hydrogen ratio ($\Sigma_{\rm H_{2}}/\Sigma_{\rm HI}$) and the surface densities of the dust (left panel), gas (middle panel), and stars (right panel). To investigate the connections between the molecular-to-neutral hydrogen ratio and star formation processes on kiloparsec scales, we bin the pixels on these diagrams and calculate the average of the perpendicular distances from the best-fit of the ensemble rSFMS relation (left panel of Figure~\ref{fig:merge_SF_bin_pixs}) of the member pixels in each bin. Then, we color code the bins based on this distance. We also show the density contours for the pixel distributions on the diagrams. The figure shows that there is an overall strong correlation between the $\text{H}_{2}$-to-\HI{} ratio and the surface densities of dust, gas, and stars, as indicated by the small scatter ($\sigma \leqslant 0.2$ dex) and high $\rho$ value ($\gtrsim 0.5$).

From these relations, we try to infer the typical ISM surface densities at the transition between the molecular and atomic hydrogen by collecting pixels that have $\Sigma_{\rm H_{2}}$ within $0.8$--$1.2 \Sigma_{\rm HI}$ (represented by the horizontal bands in Figure~\ref{fig:mol_to_neutral_ratio}), and calculate the median values of $\Sigma_{\rm dust}$, $\Sigma_{\rm gas}$, and $\Sigma_{*}$. The typical ISM surface densities at the \HI{}-to-$\text{H}_{2}$ transition are found to be: $\log(\Sigma_{\rm dust} [M_{\odot}\text{ kpc}^{-2}])=5.34\pm 0.02$, $\log(\Sigma_{\rm gas} [M_{\odot}\text{ kpc}^{-2}])= 7.15\pm 0.02$, and $\log(\Sigma_{*} [M_{\odot}\text{ kpc}^{-2}])= 7.73\pm 0.02$. These critical values are shown by the vertical dashed lines in Figure~\ref{fig:mol_to_neutral_ratio}. The uncertainties of the medians are estimated using the bootstrap resampling method, which is relatively conservative. While the uncertainties of the medians are small, the standard deviations are modest, but still considerably tight, with $0.23$, $0.17$, and $0.27$ dex for the surface densities of dust, gas, and stars, respectively. In agreement with this result, a similar study by \citet{2008Leroy} found critical surface densities of the stellar mass and gas at the \HI-to-$\text{H}_{2}$ transition of $\log(\Sigma_{*} [M_{\odot}\text{ kpc}^{-2}])=7.91$ and $\log(\Sigma_{\rm gas} [M_{\odot}\text{ kpc}^{-2}])=7.15$. 

In the rightmost panel of Figure~\ref{fig:mol_to_neutral_ratio}, the color-coding shows a clear correlation between the scatter of the $\Sigma_{*}$--$(\Sigma_{\rm H_{2}}/\Sigma_{\rm HI})$ relation and the rSFMS normalization. A clear bimodality of the rSFMS normalization is also seen on the $\Sigma_{*}$--($\Sigma_{\rm H_{2}}/\Sigma_{\rm HI}$) diagram where the majority of the sub-galactic regions on the upper envelope of this relation are those that are located on the upper envelope of the rSFMS relation and vice versa. However, we do not see a similar trend in the other two relations, where the quiescent spatial regions (rSFMS normalization $\lesssim -0.5$ dex) are mostly residing on the uppermost and lowermost envelopes of both relations, while the star-forming spatial regions (rSFMS normalization $\gtrsim 0.0$ dex) are residing around the best-fit line. In contrast to our result, \citet{2020Morselli} found a steadily decreasing $\text{H}_{2}$-to-\HI{} ratio with increasing rSFMS normalization. Unfortunately, they only analyzed the correlations between the rSFMS normalization and the scatter of the $\Sigma_{\rm gas}$--($\Sigma_{\rm H_{2}}/\Sigma_{\rm HI}$) relation. Therefore, we can not compare our results on the $\Sigma_{\rm dust}$--($\Sigma_{\rm H_{2}}/\Sigma_{\rm HI}$) and $\Sigma_{*}$--($\Sigma_{\rm H_{2}}/\Sigma_{\rm HI}$) relations with their results. This discrepancy can be caused by a possible discrepancy in the SFR estimate (which have a direct effect on the rSFMS normalization): \citet{2020Morselli} derived the SFR using a prescription involving the total UV and IR luminosities, while we have shown in \citetalias{2021Abdurrouf3} that our SFR estimates deviate from this kind of prescription mainly because of a significant contribution of old stellar populations to the dust heating. 

The small scatter ($\sigma=0.19$ dex) and high $\rho$ ($0.78$) of the $\Sigma_{*}$--($\Sigma_{\rm H_{2}}/\Sigma_{\rm HI}$) relation suggest that the surface density of stellar mass plays the strongest role in defining the $\text{H}_{2}$-to-\HI{} ratio than the other baryonic components. It also sets the color gradient clearly, such that the $\text{H}_{2}$-to-\HI{} ratio appears to be increasing with increasing rSFMS normalization at a given $\Sigma_{*}$. This may indicate the importance of the gravitational potential of the stars in governing the physical processes in the ISM which include the condensation of gas to form $\text{H}_{2}$.

The inverted trend of rSFMS normalization between the regions with low and high surface densities of dust and gas (as shown in the left and middle panels of Figure~\ref{fig:mol_to_neutral_ratio}) seems to agree with the picture in which the $\text{H}_{2}$-to-\HI{} ratio is controlled by the balance between $\text{H}_{2}$ formation in the dense environment and $\text{H}_{2}$ destruction by photo-dissociation due to the stellar feedback that is expected to dominate in the less dense environment \citep[e.g.,][]{2014Sternberg, 2020Tacconi}. Overall, a plausible explanation could be that this inverted rSFMS normalization trend is also influenced by the gravitational potential. In regions with low surface densities, the gravitational potential is weak enough for radiative feedback from the stars, particularly those of young and massive ones, to disrupt and dissociate $\text{H}_{2}$ in the regions of intense star formations \citep{2014Sternberg}. This makes the $\text{H}_{2}$ destruction dominates over the $\text{H}_{2}$ formation in more actively star-forming regions. In high density regions, the gravitational potential is strong such that stellar feedback is not strong enough to disrupt and dissociate $\text{H}_{2}$. Because of this, the $\text{H}_{2}$ destruction in star-forming regions becomes ineffective and as a result we see an increasing molecular-to-atomic gas ratio with the increasing rSFMS normalization. This trend is more prominent in the relationship between the $\text{H}_{2}$-to-\HI{} ratio and $\Sigma_{\rm gas}$, where an opposite rSFMS normalization trend is clear between the regions with surface density higher and lower than $\Sigma_{\rm gas}\sim 10^{7.8}$ $M_{\odot}\text{ kpc}^{-2}$.

\subsection{Comparison with Previous Studies on the Spatially Resolved Star Formation Scaling Relations}
\label{sec:comp_relations_prev_studies}

Having established the spatially resolved star formation scaling relations from the panchromatic SED fitting with the energy balance approach, here we compare the relations obtained from our work and those reported by previous studies in the literature, particularly those involving integral field spectroscopy (IFS) observations and determining the spatially resolved SFR from the H$\alpha$ emission that is corrected for the dust attenuation based on the Balmer decrement. We choose the following studies (and the associated surveys) as references: \citet[][]{2016CanoDiaz} (Calar Alto Legacy Integral Field Area; CALIFA; \citealt{2012Sanchez}), \citet{2018Medling} (Sydney Australian Astronomical Observatory Multi-object IFS; SAMI; \citealt{2012Croom}), \citet{2019Lin} and \citet{2021Ellison} (ALMaQUEST; \citealt{2020Lin}), and \citet{2021Pessa} (Physics at High Angular Resolution in Nearby Galaxies; PHANGS; \citealt{2021Leroy}, \citealt{2021Emsellem}). For simplicity, we do not include previous studies that used other methods for the SFR determination. This comparison
can also serve as an indirect validation for our SFR estimate.

\begin{figure*}
\centering
\includegraphics[width=0.32\textwidth]{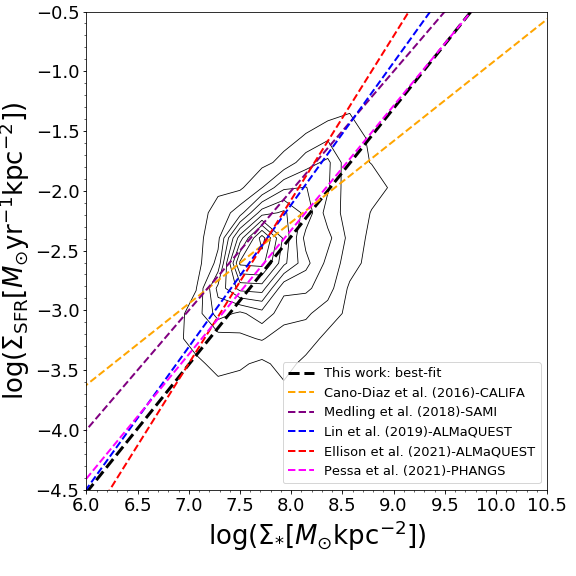}
\includegraphics[width=0.32\textwidth]{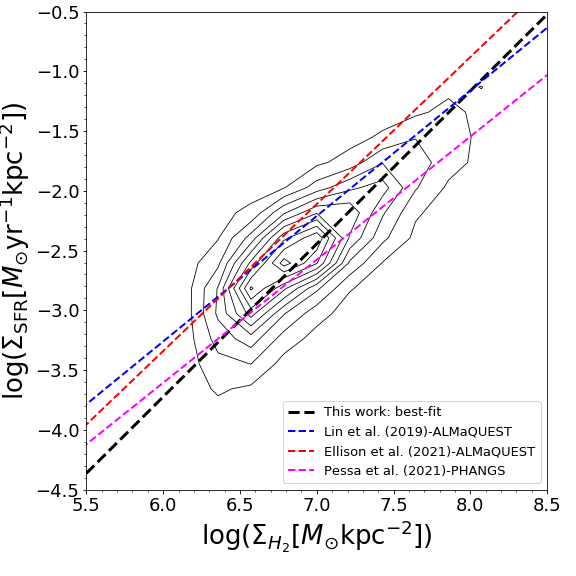}
\includegraphics[width=0.32\textwidth]{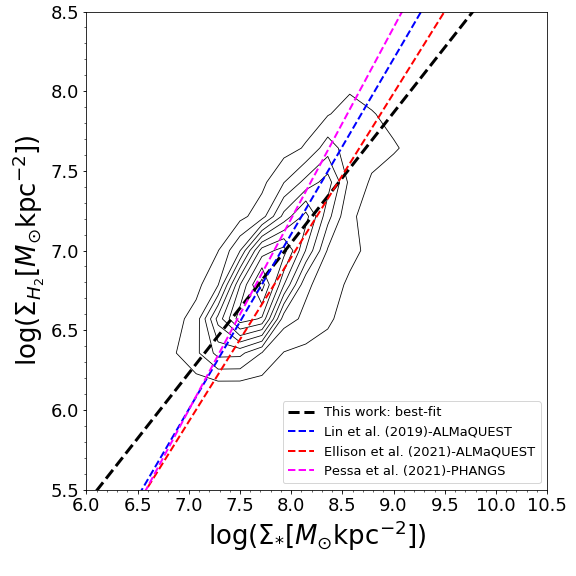}
\caption{Comparisons of the spatially resolved star formation scaling relations obtained from our work (shown by the contours and the black dashed lines) and those reported by previous studies in the literature. We only select previous studies that used IFS observations and derived the spatially resolved SFR using $H_{\alpha}$ emission that has been corrected from dust attenuation based on the Balmer decrement. Those studies include \citet{2016CanoDiaz}, \citet{2018Medling}, \citet{2019Lin}, \citet{2021Ellison}, and \citet{2021Pessa}. Overall, the scaling relations obtained from our work are consistent with those derived by the above studies.}
\label{fig:comp_relations_literature}
\end{figure*}

Figure~\ref{fig:comp_relations_literature} shows a comparison of the resolved star formation scaling relations between our work and those from the previous studies mentioned above. The relations from our work are represented by the contours and the best-fit lines (shown as the black dashed lines)  taken from Figure~\ref{fig:merge_SF_bin_pixs}. Among the aforementioned surveys, only ALMaQUEST and PHANGS surveys analyzed all  three scaling relations, while the CALIFA and SAMI only examined the rSFMS relation. The ALMaQUEST survey combines the optical IFS data from the Mapping nearby Galaxies at Apache Point Observatory \citep[MaNGA;][]{2015Bundy} survey and the CO $J=1\rightarrow 0$ emission line maps from new observations with ALMA, while the PHANGS survey combines the optical IFS data from the PHANGS-MUSE \citep{2021Emsellem} survey and the CO $J=2\rightarrow 1$ emission line maps from the PHANGS-ALMA \citep{2021Leroy} survey. We include two results from the ALMaQUEST survey because both studies analyzed different set of sample galaxies: \citet{2019Lin} analyzed only star-forming galaxies, while \citet{2021Ellison} included star-forming and green-valley galaxies in their analysis. For the PHANGS result from \citet{2021Pessa}, we only adopt relations that were derived with a spatial resolution of $1$ kpc to match that achieved in our study.

Overall, our scaling relations are consistent with those from the previous studies mentioned above. All the scaling relations seem to occupy the similar loci. The slight deviation on slope can be caused in part by the difference in sample selection.

\subsection{Interpreting the Scaling Relations at Various Scales}
\label{sec:interpret_scaling_relations}

As pointed out in Section~\ref{sec:intro}, the scaling relations among physical properties of galaxies have provided valuable insights into our current understanding of galaxy formation and evolution. Some scaling relations are intuitively understandable, such as the KS relation which reflects the star formation process from the molecular gas in the ISM, but other relations are more difficult to understand. The empirical scaling relations have provided valuable constraints for the cosmological simulations of galaxy formation \citep[e.g.,][]{2015Sparre, 2016Tacchella, 2016Taylor, 2019Donnari, 2019Trayford}. More fundamental relations like the KS can be used for fine-tuning or calibrating the sub-grid physics in the cosmological simulations, while other more ``high level'' relations (e.g.,~the SFMS and stellar mass--metallicity) can serve as a standard reference against which the simulation results will be compared (i.e.,~tested). This mutually-benefiting interactions between observations and simulations allow us to gain deeper insights into the physical mechanisms that drive the physical processes in galaxy evolution \citep[e.g.,][]{2016Tacchella, 2019Matthee, 2019Trayford}.

There have been several attempts to reproduce the global SFMS relation from cosmological simulations \citep[e.g.,][]{2016Tacchella, 2019Donnari, 2019Matthee}. \citet{2016Tacchella} used the VELA cosmological simulation \citep{2014Ceverino, 2015Zolotov} to study how star-forming galaxies evolve on the global SFMS plane across cosmic time. Despite the tightness ($\sim 0.3$ dex) of the bulk distribution of galaxies on the SFMS plane, they found that star-forming galaxies oscillate across the SFMS locus during their evolution. This confinement of galaxies in the tight sequence is regulated by the processes of gas compaction (e.g.,~by mergers, counter-rotating streams, and violent disk instabilities), depletion due to the star formation, and gas outflows. Similar attempts have also been made on local kiloparsec scales within galaxies \citep[e.g.,][]{2019Trayford, 2021Nelson}. By using the EAGLE  cosmological zoom-in hydrodynamical simulations \citep{2015Crain, 2015Schaye}, \citet{2019Trayford} studied evolution of the rSFMS and $\Sigma_{*}$--gas-phase metallicity relations. They found that the rSFMS relation started with a steep slope at high redshift ($z\sim 2$) and became shallower with time, a behavior attributed to the inside-out quenching process in galaxies. The mass--metallicity relation also tends to become flatten (at the high surface density regions) with cosmic time. They found that the AGN feedback plays a significant role in the inside-out quenching and the flattening of those two relations. 

The significant galaxy-to-galaxy variations of the resolved scaling relations presented in Section~\ref{sec:indiv_relations} indicates that those relations do not hold universally for all galaxies. Figure~\ref{fig:edit_SMs_sSFR_delta_scaling} further shows that the global sSFR and concentration index correlate strongly with the normalization of the resolved star formation scaling relations. The trend with global sSFR indicates that the relative normalization of the global SFMS among the galaxies is preserved on kiloparsec scales. On the other hand, the internal morphological factor, in particular the bulge fraction, also plays an important role in the quenching of star formation in galaxies, as has been shown here and previous studies \citep[e.g.,][]{2009Martig, 2014Genzel}.

Recently, some papers have discussed which of the resolved star formation scaling relations is more fundamental than the others \citep[e.g.,][]{2019Lin, 2020Morselli, 2021Ellison, 2021Pessa}. Considering the scatters in those relations, our results suggest that the rKS and rMGMS relations are likely independent of each other and are more fundamental from which the rSFMS relation emerges, in agreement with the results from previous studies \citep[e.g.,][]{2019Lin, 2021Ellison}. The lack of correlation between the rKS and rMGMS relations is further supported by the lack of correlation between the SFE and $f_{\rm H_{2}}$ on kiloparsec scales as shown in the right panel of Figure~\ref{fig:fH2_SFE_sSFR_indiv}. Though, we see a strong positive correlation between the normalization of the rKS and rMGMS relations for individual galaxies in Figure~\ref{fig:dSFMS_dKS_dMGMS_comp}. This may indicate that the SFE and $\text{H}_{2}$ fraction are correlated on global scales but they are somehow uncorrelated on kiloparsec scales. The constant SFE over more than an order of magnitude variation in the $\text{H}_{2}$ fraction on kiloparsec scales (the right panel of Figure~\ref{fig:fH2_SFE_sSFR_indiv}) suggests that the SFE is likely governed by physical processes that act globally. In line with this result, previous studies have shown that the SFE (and $\text{H}_{2}$ depletion time) is correlated with other global physical properties, including the sSFR, $M_{*}$, concentration index, and dynamical time, which is the timescale for the formation of stars per galactic orbital time due to the gravitational instability of the cold gas in the galactic disk \citep[e.g.,][]{1997Silk, 1998Kennicutt, 2010Genzel, 2011Saintonge, 2014Huang, 2018Tacconi}.

We have also analyzed the sSFR--SFE and $f_{\text{H}_{2}}$--sSFR relations on kiloparsec scales (left and middle panels of Figure~\ref{fig:fH2_SFE_sSFR_indiv}). Despite the tightness of these relations, we still find galaxy-to-galaxy variations that appear to separate spatial regions of different galaxies diagonally, in contrast to the vertical (i.e.,~normalization) variations we see in the other relations. It is also clear that the galaxy-to-galaxy variations in the above two relations are mainly driven by the global sSFR. The sSFR is a key parameter that describes the rate of the stellar mass growth due to star formation and it is thought to be primarily driven by the accretion history of dark matter halo, as discussed in previous studies \citep[e.g.,][]{2013Lilly, 2014Peng, 2014Dekel, 2021Duo}. It has also been suggested that the sSFR and SFE are key fundamental parameters in galaxies \citep[e.g.,][]{2014Huang, 2021Duo}. Both of these parameters can be considered as the primary parameters for the evolution of the stellar populations and gas in galaxies \citep{2010Peng, 2020Tacconi}.

While the resolved star formation scaling relations have been studied extensively in the last decade, the resolved scaling relations between the dust surface density and other physical parameters are still unexplored. We find tight correlations between the surface densities of dust and gas (in both molecular and total gas) which applies universally to all galaxies in our sample, as indicated by the lack of the galaxy-to-galaxy variation in these relations (see Figure~\ref{fig:edit_Mdust_MHI_MH2_SM_pix_indiv}). 
Despite the variation in global properties (e.g.,~sSFR, SFE, $M_{*}$, and $C$), the sub-galactic regions in the galaxies follow these tight relations. The tightness and the universality of the $\Sigma_{\rm dust}$--$\Sigma_{\rm H_{2}}$ relation supports the view that dust plays important roles in catalyzing the formation of molecular gas \citep[e.g.,][]{1979Hollenbach, 2011Yamasawa}. Moreover, as demonstrated in Figure~\ref{fig:dSF_ddust_comp}, galaxies with higher levels of resolved dust-to-stellar mass ratios tend to have higher levels of resolved sSFR, SFE, and $\text{H}_{2}$ fraction. This is consistent with the picture that dust also plays an important role in promoting the star formation (e.g., by shielding the gas from the ISRF which can cool the gas; \citealt{1999Hollenbach, 2009Krumholz, 2011Yamasawa, 2012Glover}).

\section{Summary and Conclusions} 
\label{sec:conclusion}

We investigate the scaling relations on kiloparsec scales by exploiting high resolution maps of the spatially resolved properties, including the surface densities of stars, SFR, dust, and the atomic and molecular gas, of ten nearby spiral galaxies that were obtained in \citet{2021Abdurrouf3}. The surface densities of stars ($\Sigma_{*}$), dust ($\Sigma_{\rm dust}$), and SFR ($\Sigma_{\rm SFR}$) were obtained by performing the spatially resolved SED fitting on FUV--FIR imaging data using the \verb|piXedfit| software, while the surface densities of the atomic and molecular gas ($\Sigma_{\rm HI}$ and $\Sigma_{\rm H_{2}}$) were obtained from the archival data. We analyze two sets of spatially resolved scaling relations in this work: the star formation scaling relations ($\Sigma_{\rm H_{2}}$--$\Sigma_{\rm SFR}$--$\Sigma_{*}$) and dust scaling relations (between $\Sigma_{\rm dust}$ and the other properties). 

Our key results are summarized below: 
\begin{enumerate}
\item The resolved star formation scaling relations formed with ensemble of all spatial regions of the sample galaxies are reasonably tight (Figure~\ref{fig:merge_SF_bin_pixs}). The ensemble resolved star-forming main sequence (rSFMS) has larger scatter ($0.31$ dex) than the other two relations (resolved Kennicutt--Schmidt and molecular gas main sequence; rKS and rMGMS), which interestingly have equal scatter ($0.19$ dex) and Spearman rank-order correlation coefficient ($\rho=0.76$). The slopes of the ensemble rSFMS, rKS, and rMGMS are linear, super-linear, and sub-linear, respectively, which suggest that the sub-galactic regions in the galaxies have on average broadly constant rate of the stellar mass growth by star formation, increasing SFE with increasing $\Sigma_{\rm H_{2}}$, and decreasing $\text{H}_{2}$ fraction ($f_{\rm H_{2}}$) with increasing $\Sigma_{*}$, respectively.

\item We observe tight, positive correlations between the dust surface density ($\Sigma_{\rm dust}$) and the surface densities of gas (both molecular and atomic), stars, and SFR (Figure~\ref{fig:edit_Mdust_MHI_MH2_bins}). The $\Sigma_{\rm dust}$--$\Sigma_{\rm gas}$ relation is the tightest ensemble relation we examine in this work. We also find that $\Sigma_{\rm dust}$ correlates more strongly with $\Sigma_{\rm H_{2}}$ than with $\Sigma_{\rm HI}$. On the other hand, we find no significant correlation between $\Sigma_{\rm dust}$ and sSFR.

\item We find that the majority of the scaling relations exhibit significant galaxy-to-galaxy variations in terms of the normalization, slope, and shape, indicating that they do not apply universally to the sub-galactic regions of all the galaxies (Figures~\ref{fig:comb_SFMS_KS_MGMS_indiv_gals} and~\ref{fig:edit_Mdust_MHI_MH2_SM_pix_indiv}). Among the scaling relations analyzed in this work, only $\Sigma_{\rm dust}$--$\Sigma_{\rm H_{2}}$ and $\Sigma_{\rm dust}$--$\Sigma_{\rm gas}$ exhibit no significant galaxy-to-galaxy variation (Figure~\ref{fig:edit_Mdust_MHI_MH2_SM_pix_indiv}), which suggests that they hold similarly among our sample galaxies.

\item We find a clear connection between the normalization of the global SFMS ($\Delta$SFMS) and the normalization of the resolved scaling relations. For the rSFMS, rKS, rMGMS, and $\Sigma_{\rm dust}$--$\Sigma_{\rm SFR}$ relations, sub-galactic regions of the galaxies with higher $\Delta$SFMS tend to have higher normalization in the resolved scaling relations, while an opposite trend is found in the $\Sigma_{\rm dust}$--$\Sigma_{*}$ relation (Figures~\ref{fig:comb_SFMS_KS_MGMS_indiv_gals} and~\ref{fig:edit_Mdust_MHI_MH2_SM_pix_indiv}). Related to this, an interesting trend is shown in the resolved $f_{\rm H_{2}}$--sSFR and SFE--sSFR relations, where the sub-galactic regions of the galaxies with higher (lower) $\Delta$SFMS tend to reside in the top right (bottom left) side (i.e.,~sub-galactic regions of different galaxies are separated diagonally; Figure~\ref{fig:fH2_SFE_sSFR_indiv}).

\item We find significant correlations among the normalization offsets of the resolved star formation scaling relations ($\Delta$rSFMS, $\Delta$rKS, and $\Delta$rMGMS) which suggest that the galaxies with higher levels of $\text{H}_{2}$ fraction tend to have higher SFE and sSFR in their sub-galactic regions (Figure~\ref{fig:dSFMS_dKS_dMGMS_comp}). We also find correlations between the normalization offsets of the dust scaling relations ($\Sigma_{\rm dust}$--$\Sigma_{*}$ and $\Sigma_{\rm dust}$--$\Sigma_{\rm SFR}$) and the star formation scaling relations, which overall suggest that galaxies with higher levels of resolved dust-to-stellar mass ratios tend to have higher levels of resolved sSFR, SFE, and molecular gas fraction (Figure~\ref{fig:dSF_ddust_comp}).

\item We find strong positive (negative) correlations between the normalization offsets of the resolved star formation scaling relations and the global sSFR (concentration index; $C$), while only a weak correlation is found with the global $M_{*}$ (Figure~\ref{fig:edit_SMs_sSFR_delta_scaling}). The correlations with global sSFR may indicate that the global processes play important roles in governing the star formation activities in galaxies, while the correlation with $C$ argues for significant effects of internal processes related to the central bulge component. We interpret the latter as evidence for the morphological quenching in galaxies.

\item We observe tight relationships between the molecular-to-atomic gas ratio and the surface densities of stars, gas, and dust (Figure~\ref{fig:mol_to_neutral_ratio}). From these relations, we find that the typical ISM surface densities at the molecular-to-atomic transition are: $\log(\Sigma_{\rm dust} [M_{\odot}\text{ kpc}^{-2}])=5.34\pm 0.02$, $\log(\Sigma_{\rm gas} [M_{\odot}\text{ kpc}^{-2}])= 7.15\pm 0.02$, and $\log(\Sigma_{*} [M_{\odot}\text{ kpc}^{-2}])= 7.73\pm 0.02$. In the relation with $\Sigma_{*}$, we see a clear trend of increasing rSFMS normalization with increasing molecular-to-atomic gas ratio at fixed $\Sigma_{*}$. We see a similar trend of the rSFMS normalization in the $\Sigma_{\rm gas}$--($\Sigma_{\rm H_{2}}/\Sigma_{\rm HI}$) relation at $\Sigma_{\rm gas}\gtrsim 10^{7.8} M_{\odot}\text{ kpc}^{-2}$ but the trend is inverted at the lower gas surface density. We find that the $\Sigma_{*}$--($\Sigma_{\rm H_{2}}/\Sigma_{\rm HI}$) is the tightest among the three relations, which imply the importance of local gravitational potential in governing the physical processes in the ISM, including the condensation of gas into molecular clouds. We also note the possibility of $\text{H}_{2}$ destruction due to photo-dissociation in star-forming regions with relatively low surface densities.
\end{enumerate}

Currently, the spatially resolved panchromatic SED fitting analysis such as the one carried out in this study is only limited for nearby galaxies because of the lack of imaging data in the infrared that have sufficiently high spatial resolution. With the launch of the James Webb Space Telescope with its depth and high spatial resolution imaging in the infrared regime, combined with the optical and near-infrared images from the Hubble Space Telescope, we can push this analysis toward intermediate redshifts. It is also possible to combine it with the resolved CO map from the radio interferometer, such as the Atacama Large Millimeter/submillimeter Array (ALMA), to resolve molecular gas distribution in the galaxies. With this we can study evolution of the spatially resolved properties of the stellar population, dust, and gas in galaxies.

%% Also note that the akcnowlodgment environment does not support long amounts of text. If you have a lot of people and institutions to acknowledge, do not use this command. Instead, create a new

\section{Acknowledgments}.
%\begin{acknowledgments}
We are grateful for support from the Ministry of Science and Technology of Taiwan under grants MOST 109-2112-M-001-005 and MOST 110-2112-M-001-004, and a Career Development Award from Academia Sinica (AS-CDA-106-M01). H.H. thanks the Ministry of Science and Technology of Taiwan for support through grant MOST 108-2112-M-001-007-MY3, and the Academia Sinica for Investigator Award AS-IA-109-M02. P.F.W. acknowledges the support of the fellowship from the East Asian Core Observatories Association. The computations in this research were run on the TIARA cluster at ASIAA.

This work is based on observations made with the NASA Galaxy Evolution Explorer (GALEX), which is operated for NASA by the California Institute of Technology under NASA contract NAS5-98034.

Funding for the Sloan Digital Sky Survey IV has been provided by the Alfred P. Sloan Foundation, the U.S. Department of Energy Office of Science, and the Participating Institutions. SDSS-IV acknowledges support and resources from the Center for High-Performance Computing at the University of Utah. The SDSS web site is www.sdss.org. SDSS-IV is managed by the Astrophysical Research Consortium for the Participating Institutions of the SDSS Collaboration including the Brazilian Participation Group, the Carnegie Institution for Science, Carnegie Mellon University, the Chilean Participation Group, the French Participation Group, Harvard-Smithsonian Center for Astrophysics, Instituto de Astrof\'isica de Canarias, The Johns Hopkins University, Kavli Institute for the Physics and Mathematics of the Universe (IPMU)/University of Tokyo, the Korean Participation Group, Lawrence Berkeley National Laboratory,  Leibniz Institut f\"ur Astrophysik Potsdam (AIP), Max-Planck-Institut f\"ur Astronomie (MPIA Heidelberg), Max-Planck-Institut f\"ur Astrophysik (MPA Garching), Max-Planck-Institut f\"ur Extraterrestrische Physik (MPE), National Astronomical Observatories of China, New Mexico State University, New York University, University of Notre Dame, Observat\'ario Nacional / MCTI, The Ohio State University, Pennsylvania State University, Shanghai Astronomical Observatory, United Kingdom Participation Group, Universidad Nacional Aut\'onoma de M\'exico, University of Arizona, University of Colorado Boulder, University of Oxford, University of Portsmouth, University of Utah, University of Virginia, University of Washington, University of Wisconsin, Vanderbilt University, and Yale University. 

This publication makes use of data products from the Two Micron All Sky Survey, which is a joint project of the University of Massachusetts and the Infrared Processing and Analysis Center/California Institute of Technology, funded by the National Aeronautics and Space Administration and the National Science Foundation. 

This publication makes use of data products from the Wide-field Infrared Survey Explorer, which is a joint project of the University of California, Los Angeles, and the Jet Propulsion Laboratory/California Institute of Technology, funded by the National Aeronautics and Space Administration. 

This work is based in part on observations made with the Spitzer Space Telescope, which was operated by the Jet Propulsion Laboratory, California Institute of Technology under a contract with NASA. 

Herschel is an ESA space observatory with science instruments provided by European-led Principal Investigator consortia and with important participation from NASA.
%\end{acknowledgments}


\vspace{5mm}
\facilities{GALEX, Sloan, FLWO:2MASS, WISE, \textit{Spitzer}, \textit{Herschel}, IRAM:30m, VLA}


\software{
\texttt{piXedfit} \citep{2021Abdurrouf2, 2021Abdurrouf},
\texttt{Astropy}  \citep{2013Astropy, 2018Astropy},
\texttt{SExtractor}  \citep{1996bertin},
\texttt{FSPS}  \citep{2009Conroy,2010Conroy},
\texttt{emcee}  \citep{2013Foreman},
\texttt{Photutils}  \citep{2019Bradley},
\texttt{matplotlib} \citep{2007Hunter}
}



\appendix

\section{Appendix information}

\section{Effect of the Metallicity-dependent CO-to-$\rm{H}_{2}$ Conversion Factor} \label{sec:effect_varian_alpha_CO}

Our analysis of the molecular gas is subject to the uncertainty in the assumption of the CO-to-$\text{H}_{2}$ conversion factor ($\alpha_{\rm CO}$). As we have shown in \citetalias{2021Abdurrouf3} (Section 5.5 therein), the $\alpha_{\rm CO}$ assumption gives a systematic effect on the radial profiles (or gradients) of the quantities that depend on the $\text{H}_{2}$ abundance. Therefore, it is important to check the effect of the $\alpha_{\rm CO}$ assumption on the scaling relations that we analyze in this paper, which are derived based on the assumption of a constant $\alpha_{\rm CO}$ (see Section~\ref{sec:data_analysis}). As we have used the $\alpha_{\rm CO}$ prescription of \citet{2012Schruba} for examining the effect of the metallicity-dependent $\alpha_{\rm CO}$ in \citetalias{2021Abdurrouf3}, we use it again here. We refer the reader to \citetalias{2021Abdurrouf3} for the metallicity-dependent $\alpha_{\rm CO}$ prescription of \citet{2012Schruba} and how to implement it. It is important to note that there is an inherent limitation in applying the metallicity-dependent $\alpha_{\rm CO}$ for deriving the spatially resolved scaling relations in our analysis because we only have the gas-phase metallicity ($12+\log(\text{O}/\text{H})$) in the form of a radial gradient, without the azimuthal variations. In our analysis, pixels that are located at the same elliptical semi-major distance from the galactic center have the same $\alpha_{\rm CO}$. In this section, we re-calculate the spatially resolved scaling relations that directly or indirectly involve $\text{H}_{2}$, including rKS, rMGMS, $f_{\rm H_{2}}$--SFE, $f_{\rm H_{2}}$--sSFR, $\Sigma_{\rm dust}$--$\Sigma_{\rm H_{2}}$, and $\Sigma_{\rm dust}$--$\Sigma_{\rm gas}$. Moreover, we also check the effects of metallicity-dependent $\alpha_{\rm CO}$ on the relationships between the $\text{H}_{2}$-to-\HI{} ratio and the surface densities of dust, gas, and stars. Here, we only analyze six galaxies in our sample that have data of \HI{}, $\text{H}_{2}$, and $12+\log(\text{O}/\text{H})$ gradient. We adopt the data of gas-phase metallicity for NGC 628, NGC 3184, NGC 5194, and NGC 5457 from \citet{2020Berg} and for NGC 3351 and NGC 5055 from \citet{2014Pilyugin}. \citet{2014Pilyugin} measured $12+\log(\text{O}/\text{H})$ based on the analysis of the strong emission lines ([\ion{O}{II}] $\lambda 3727+\lambda 3729$, [\ion{O}{III}] $\lambda 5007$, [\ion{N}{II}] $\lambda 6584$, and [\ion{S}{II}] $\lambda 6717+ \lambda 6731$), while \citet{2020Berg}, which is part of the CHemical Abundances Of Spirals survey \citep[CHAOS;][]{2015Berg}, uses the so-called ``direct method'' based on the measurement of the electron temperature from auroral lines (e.g.,~[\ion{S}{II}] $\lambda 4068+\lambda 4076$, [\ion{O}{III}] $\lambda 4363$, [\ion{N}{II}] $\lambda 5755$, [\ion{S}{III}] $\lambda 6312$, [\ion{O}{II}] $\lambda 7320+\lambda 7330$). The latter method is considered to be the most accurate technique for measuring gas-phase metallicity. However, until now, only metallicity gradient of four galaxies are published by the CHAOS survey, all of which used in our analysis.

\begin{figure*}
\centering
\includegraphics[width=1.0\textwidth]{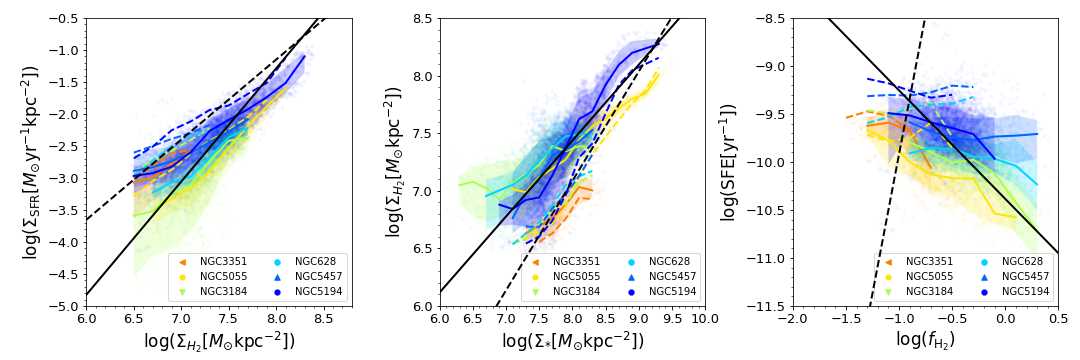} \includegraphics[width=1.0\textwidth]{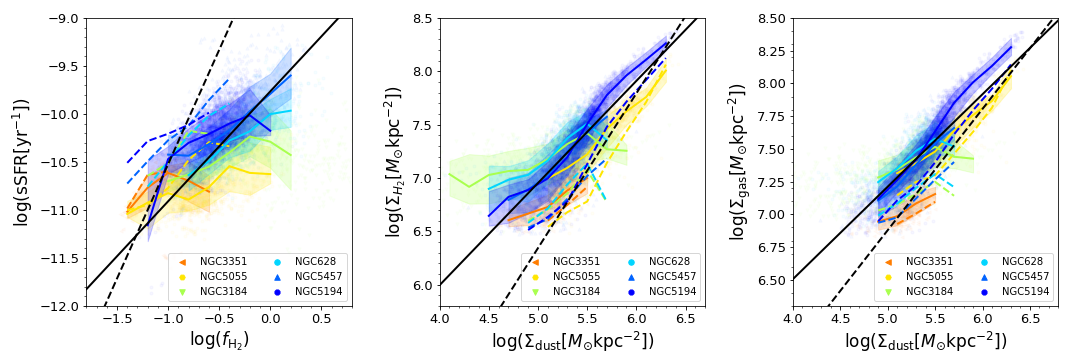}
\caption{Comparisons between the spatially resolved scaling relations obtained based on the assumption of constant $\alpha_{\rm CO}$ and metallicity-dependent $\alpha_{\rm CO}$ \citep{2012Schruba}. The scaling relations that are re-analyzed here are those that directly or indirectly involve $\text{H}_{2}$, including rKS, rMGMS, $f_{\rm H_{2}}$--SFE, $f_{\rm H_{2}}$--sSFR, $\Sigma_{\rm dust}$--$\Sigma_{\rm H_{2}}$, and $\Sigma_{\rm dust}$--$\Sigma_{\rm gas}$. Only six galaxies that have data of \HI{}, $\text{H}_{2}$, and gas-phase metallicity gradient are analyzed here. The relations obtained based on the metallicity-dependent $\alpha_{\rm CO}$ assumption are represented by the colored data points. The colored solid lines and shaded areas represent the median profiles and the 16th--84th percentile ranges, respectively. The median profiles that are obtained with the constant $\alpha_{\rm CO}$ are shown by the colored dashed lines, which are the same as those in Figures~\ref{fig:comb_SFMS_KS_MGMS_indiv_gals},~\ref{fig:fH2_SFE_sSFR_indiv}, and~\ref{fig:edit_Mdust_MHI_MH2_SM_pix_indiv}. The solid and dashed linear lines are the best-fit to the ensemble of those scaling relations for the six galaxies with metallicity-dependent and constant $\alpha_{\rm CO}$, respectively.}
\label{fig:pixs_effect_alphaCO_S12}
\end{figure*}

Figure~\ref{fig:pixs_effect_alphaCO_S12} shows comparisons between the scaling relations of individual galaxies that are obtained based on the assumption of the constant $\alpha_{\rm CO}$ (dashed line profiles, which are the same as those in Figures~\ref{fig:comb_SFMS_KS_MGMS_indiv_gals},~~\ref{fig:edit_Mdust_MHI_MH2_SM_pix_indiv}, and~\ref{fig:fH2_SFE_sSFR_indiv}) and those obtained based on the metallicity-dependent $\alpha_{\rm CO}$ (data points and solid line profiles with shaded area, which represent 50th and 16th--84th percentiles). We fit the resolved star formation and dust scaling relations from the ensemble data of the six galaxies and summarize the results in Table~\ref{tab:compare_alphaCO}. In Figure~\ref{fig:pixs_effect_alphaCO_S12}, we also show the best-fit linear relations with the metallicity-dependent and constant $\alpha_{\rm CO}$ (solid and dashed lines, respectively). Since these dashed lines are not derived from the full sample, they are different from those presented in Figures~\ref{fig:comb_SFMS_KS_MGMS_indiv_gals},~~\ref{fig:edit_Mdust_MHI_MH2_SM_pix_indiv}, and~\ref{fig:fH2_SFE_sSFR_indiv}. As can be seen, the metallicity-dependent $\alpha_{\rm CO}$ changes the slope and normalization of the scaling relations (both for the individual galaxies and for the ensemble ones) in a complex way. The fitting results in Table~\ref{tab:compare_alphaCO} show that applying a metallicity-dependent $\alpha_{\rm CO}$ results in a slightly larger scatter in the scaling relations compared to the case with a constant $\alpha_{\rm CO}$, although they are still considered to be tight ($\sigma \lesssim 0.2$ dex). The increase in scatter can be caused by at least two factors: the systematic effect of the metallicity on the dust-to-gas mass ratio and the difference in the gas-phase metallicity diagnostics and calibration used. Overall, the radial metallicity gradients used in our analysis have a dynamical range of $8.3$ to $8.85$ dex. The fitting results also indicate that reducing the number of galaxies while still applying the same constant $\alpha_{\rm CO}$ assumption results in a slightly smaller scatter in the relations. Another point we can make from Figure~\ref{fig:pixs_effect_alphaCO_S12} is that the new relations derived with a metallicity-dependent $\alpha_{\rm CO}$ also exhibits galaxy-to-galaxy variations and it is thus galaxy-to-galaxy variations that significantly contribute to the scatter of the ensemble relations. This is consistent with the results we obtained with the constant $\alpha_{\rm CO}$ as discussed in Section~\ref{sec:indiv_relations}, and is thus unaffected by the assumption on $\alpha_{\rm CO}$. In comparison with the scaling relations obtained with the constant $\alpha_{\rm CO}$, the metallicity-dependent $\alpha_{\rm CO}$ produces steeper slopes and lower normalization in the rKS relation, shallower slopes and higher normalization in the rMGMS, $\Sigma_{\rm dust}$--$\Sigma_{\rm H_{2}}$, and $\Sigma_{\rm dust}$--$\Sigma_{\rm gas}$ relations, and shallower slopes and lower normalization in the $f_{\rm H_{2}}$--SFE and $f_{\rm H_{2}}$--sSFR relations. Among the relations analyzed here, the $f_{\rm H_{2}}$--SFE of individual galaxies are significantly influenced by the $\alpha_{\rm CO}$ choice. They are broadly flat when the constant $\alpha_{\rm CO}$ is assumed, while they have a negative slope when the metallicity-dependent $\alpha_{\rm CO}$ is assumed.

\begin{deluxetable}{ccccccccc}[ht]
\tablenum{4}
\tablecaption{Comparison of the Linear Function Fits Between the Scaling Relations Derived with the Constant and Metallicity-dependent $\alpha_{\rm CO}$ \label{tab:compare_alphaCO}}
\tablewidth{0pt}
\tablehead{
\colhead{} & \multicolumn{8}{c}{$\alpha_{\rm CO}$ assumption} \\
\cline{2-9}
\colhead{Correlation} & \multicolumn{4}{c}{Constant $\alpha_{\rm CO}$} &  \multicolumn{4}{c}{Metallicity-dependent $\alpha_{\rm CO}$} \\
\cline{2-9}
\colhead{} & \colhead{$\rho$} & \colhead{$\alpha$} & \colhead{$\beta$} & \colhead{$\sigma$} & \colhead{$\rho$} & \colhead{$\alpha$} & \colhead{$\beta$} & \colhead{$\sigma$}
}
\decimals
\startdata
rKS & $0.75$ & $1.25 \pm 0.01$ & $-11.15 \pm 0.10$ & $0.18$ & $0.78$ & $1.77 \pm 0.03$ & $-15.43 \pm 0.19$ & $0.20$ \\
rMGMS & $0.83$ & $0.95 \pm 0.01$ & $-0.51 \pm 0.06$ & $0.14$ & $0.71$ & $0.66 \pm 0.01$ & $2.16 \pm 0.07$ & $0.21$ \\
$\Sigma_{\rm dust}$--$\Sigma_{\rm H_{2}}$ & $0.76$ & $1.45 \pm 0.01$ & $-0.91 \pm 0.08$ & $0.13$ & $0.79$ & $0.96 \pm 0.02$ & $2.17 \pm 0.09$ & $0.18$ \\
$\Sigma_{\rm dust}$--$\Sigma_{\rm gas}$ & $0.85$ & $0.93 \pm 0.01$ & $2.24 \pm 0.05$ & $0.09$ & $0.83$ & $0.71 \pm 0.01$ & $3.68 \pm 0.07$ & $0.14$ \\
\enddata
%\tablecomments{}
\end{deluxetable}

\begin{figure*}
\centering
\includegraphics[width=0.32\textwidth]{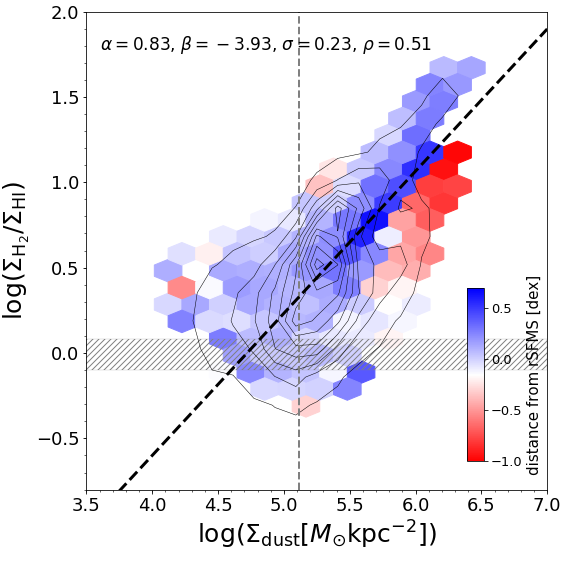}
\includegraphics[width=0.32\textwidth]{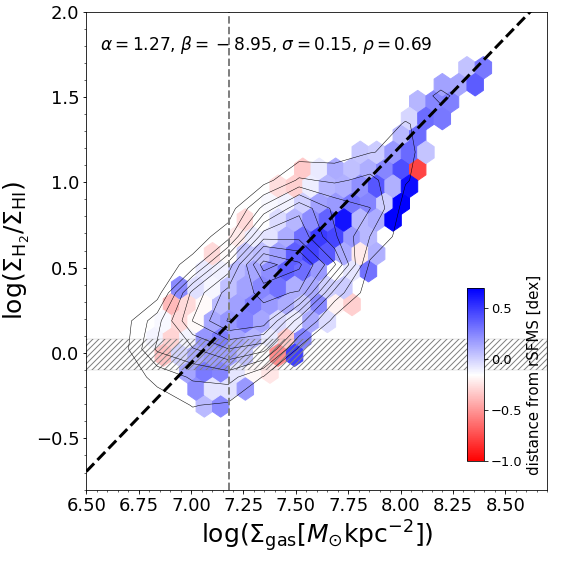}
\includegraphics[width=0.32\textwidth]{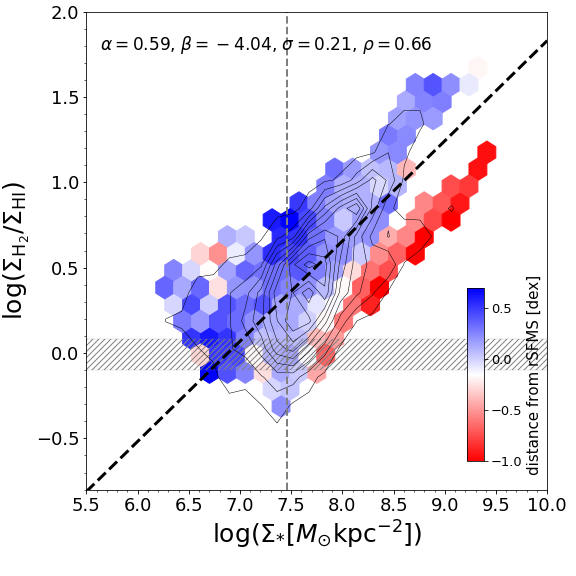}
\caption{The relationship between the $\text{H}_{2}$-to-\HI{} ratio and the surface densities of dust (left panel), gas (middle panel), and stars (right panel) that are obtained with the metallicity-dependent $\alpha_{\rm CO}$ \citep{2012Schruba}. The symbols in this figure are the same as those in Figure~\ref{fig:mol_to_neutral_ratio}.}
\label{fig:mol_to_neutral_ratio_alphaCO_S12}  
\end{figure*}

In Figure~\ref{fig:mol_to_neutral_ratio_alphaCO_S12}, we show the relations between the $\text{H}_{2}$-to-\HI{} ratio and the surface densities of dust (left panel), gas (middle panel), and stars (right panel) that are obtained with the metallicity-dependent $\alpha_{\rm CO}$. As we can see, the relations hold tightly, comparable to those derived based on the constant $\alpha_{\rm CO}$. The color-coding in Figure~\ref{fig:mol_to_neutral_ratio_alphaCO_S12}, which represents the rSFMS normalization, show consistency with the right panel of Figure~\ref{fig:mol_to_neutral_ratio}, but not with the other two panels. In the right panel, the rSFMS normalization tends to increase with increasing $\text{H}_{2}$-to-\HI{} ratio at a given $\Sigma_{*}$. In the middle panel, for regions with $\Sigma_{\rm gas} \lesssim 10^{7.75}M_{\odot}\text{ kpc}^{-2}$, the rSFMS normalization tends to decrease with increasing $\text{H}_{2}$-to-\HI{} ratio at a given $\Sigma_{\rm gas}$, which is consistent with the trend we obtain with the constant $\alpha_{\rm CO}$. However, we observe different trends between the two $\alpha_{\rm CO}$ assumptions in the higher $\Sigma_{\rm gas}$ regions, where there appear to be no $\Delta$rSFMS gradient (i.e.,~dominated by star-forming sub-galactic regions) in the result with the metallicity-dependent $\alpha_{\rm CO}$, whereas we see a trend of increasing $\Delta$rSFMS with the $\text{H}_{2}$-to-\HI{} ratio (at a given $\Sigma_{\rm gas}$) in the result with the constant $\alpha_{\rm CO}$. A different trend is also shown in the $\Sigma_{\rm dust}$--$(\Sigma_{\rm H_{2}}/\Sigma_{\rm HI})$ relation where the bimodality of $\Delta$rSFMS in the low $\Sigma_{\rm dust}$ ($\lesssim 10^{5.7} M_{\odot}\text{ kpc}^{-2}$), which appears when we use the constant $\alpha_{\rm CO}$, is not clear in the relation obtained with the metallicity-dependent $\alpha_{\rm CO}$. In the higher dust surface density regions, the quiescent regions that appear in the upper envelope when we use the constant $\alpha_{\rm CO}$ are not present when we use the metallicity-dependent $\alpha_{\rm CO}$. However, both conversion factors produce a consistent result in that the lower envelope are mostly occupied by quiescent sub-galactic regions. The consistent trend observed in the $\Sigma_{*}$--$(\Sigma_{\rm H_{2}}/\Sigma_{\rm HI})$ between the constant $\alpha_{\rm CO}$ and the metallicity-dependent one further suggests the importance of the gravitational potential of the stars in governing the ISM condition, particularly in promoting the \HI{}-to-$\text{H}_{2}$ transition. We note that the number of galaxies analyzed with the metallicity-dependent $\alpha_{\rm CO}$ and the constant one are not the same, which may in part cause the discrepancies discussed above. We also highlight again the limitation caused by the lack of spatially resolved gas-phase metallicity maps in this analysis, which may affect the distribution of the $\text{H}_{2}$-to-\HI{} ratio on the three diagrams in Figure~\ref{fig:mol_to_neutral_ratio_alphaCO_S12} and cause the discrepancy of the $\Delta$rSFMS trend observed here and the one obtained with the constant $\alpha_{\rm CO}$.      

\bibliography{sample631}{}
\bibliographystyle{aasjournal}



\end{document}